\documentclass[preprint,preprintnumbers,jcp,showpacs,superscriptaddress]{revtex4-1}

\usepackage{palatino}
\usepackage[latin1]{inputenc}
\usepackage{epsf}
\usepackage{amsmath,amssymb}
\usepackage{latexsym}
\usepackage{calc}
\usepackage{color}
\usepackage{shadow}
\usepackage{epsfig}
\usepackage[pdftex, pdfstartview = {FitH}]{hyperref}
\usepackage{graphicx}

\def\bP{{\bf P}}
\def\bA{{\bf A}}

\def\dulR{{\underline{\underline{\bf R}}}}
\def\dulk{{\underline{\underline{\bf k}}}}
\def\dulr{{\underline{\underline{\bf r}}}}
\def\dulP{{\bf P}}

\begin{document}
\title{Classical nuclear motion coupled to electronic non-adiabatic transitions}
\author{Federica Agostini}
\affiliation{Max-Planck Institut f\"ur Mikrostrukturphysik, Weinberg 2, D-06120 Halle, Germany}
\author{Ali Abedi}
\affiliation{Max-Planck Institut f\"ur Mikrostrukturphysik, Weinberg 2, D-06120 Halle, Germany}
\author{E. K. U. Gross}
\affiliation{Max-Planck Institut f\"ur Mikrostrukturphysik, Weinberg 2, D-06120 Halle, Germany}

\date{\today}
\begin{abstract}
We present a detailed derivation and numerical tests of a new mixed quantum-classical scheme to deal with non-adiabatic processes. The method is presented as the zero-th order approximation to the exact coupled dynamics of electrons and nuclei offered by the factorization of the electron-nuclear wave function [A. Abedi, N. T. Maitra and E. K. U. Gross, \textsl{Phys. Rev. Lett.} \textbf{105} (2010)]. Numerical results are presented for a model system for non-adiabatic charge transfer in order to test the performance of the method and to validate the underlying approximations.
\end{abstract}
\maketitle

\section{Introduction}\label{sec: intro}
Among the ultimate goals of condensed matter physics and theoretical chemistry is the atomistic description of phenomena such as vision~\cite{cerulloN2010, schultenBJ2009, ishidaJPCB2012}, photo-synthesis~\cite{tapaviczaPCCP2011, flemingN2005}, photo-voltaic processes~\cite{rozziNC2013, silvaNM2013, jailaubekovNM2013}, proton-transfer and hydrogen storage~\cite{sobolewski, varella, hammes-schiffer, marx}. These phenomena involve the coupled dynamics of electrons and nuclei beyond the Born-Oppenheimer (BO), or adiabatic, regime and therefore require the explicit treatment of excited states dynamics. Being the exact solution of the full dynamical problem unachievable for realistic molecular systems, as the numerical cost for solving the time-dependent Schr\"odinger equation (TDSE) scales exponentially with the number of degrees of freedom, approximations need to be introduced. Usually, a quantum-classical (QC) description of the full system is adopted, where only a small number of degrees of freedom, e.g. electrons or protons, are treated quantum mechanically, while the remaining degrees of freedom are considered as classical particles, e.g. nuclei or ions. In this context, the challenge resides in determining the force that generates the classical trajectory as effect of the quantum subsystem. As the BO approximation breaks down, this electronic quantum effect on the classical nuclei cannot be expressed by the single adiabatic potential energy surface (PES) corresponding to the occupied eigenstate of the BO Hamiltonian. An exact treatment would require to take into account several adiabatic PESs that are coupled via electronic non-adiabatic transitions in regions of strong coupling, as avoided crossings or conical intersections. In the approximate QC description, however, the concept of single PES and single force that drives the classical motion is lost. In order to provide an answer to the question ``What is the classical force that generates a classical trajectory in a quantum environment subject to non-adiabatic transitions?'', different approaches to QC non-adiabatic dynamics have been proposed for the past 50 years~\cite{ehrenfest, mclachlan, shalashilin_JCP2009, tully1990, kapral-ciccotti, io, ivano, bonellaJCP2005, marxPRL2002, martinezJPCA2000, maddox, aleksandrov, pechukas, tully1971, meyer, subotnik, herman, heller, coker, prezhdo, kosloff, antoniou, stock}, but the problem still remains a challenge.

This paper investigates an alternative point of view on this longstanding problem. The recently proposed exact factorization of the electron-nuclear wave function~\cite{AMG,AMG2} allows to decompose the coupled dynamics of electrons and nuclei such that a time-dependent vector potential and a time-dependent scalar potential generate the nuclear evolution in a Schr\"odinger-like equation. The time-dependent potentials represent the exact effect of the electrons on the nuclei, beyond the adiabatic regime. This framework offers the same advantages of the BO approximation, since the \textit{single} force that generates the classical trajectory in the QC description can be determined from the potentials. We have extensively investigated the properties of the scalar potential~\cite{steps, long_steps, long_steps_mt} in situations where it can be calculated exactly, by setting the vector potential to zero with an appropriate choice of the gauge, also in the context of the QC approximation. Here, we describe a procedure~\cite{mqc} to derive an approximation to the vector and scalar potentials, that leads to a new mixed QC (MQC) approach to the coupled non-adiabatic dynamics of electrons and nuclei. This MQC scheme is presented as zero-th order approximation to the exact electronic and nuclear equations and intends to adequately describe those situations where nuclear quantum effects associated with zero-point energy, tunnelling or interference are not relevant. It is worth noting that the factorization, and the consequent decomposition of the dynamical problem, is irrespective of the specific properties, e.g. the masses, of the two sets of particles. Therefore, it can be generalized to any two-component system.

The paper is organized as follows. In Section~\ref{sec: background} we recall the exact factorization~\cite{AMG, AMG2}. The procedure to derive the classical limit of the nuclear equation is described in Section~\ref{sec: limit}, with reference to the most common approaches to QC dynamics. The new MQC scheme is presented in Section~\ref{sec: mqc} and applied to a model system for non-adiabatic charge transfer in Section~\ref{sec: results}. Here, apart from testing the performance of the new algorithm, the hypothesis underlying the classical approximation are validated by the numerical analysis. Our conclusions are presented in Section~\ref{sec: conclusion}.

\section{Theoretical background}\label{sec: background}
In the absence of an external field, the non-relativistic Hamiltonian
\begin{equation}
 \hat H = \hat T_n+\hat H_{BO}
\end{equation}
describes a system of interacting nuclei and electrons. Here, $\hat T_n$ denotes the nuclear kinetic energy and $\hat H_{BO}(\dulr,\dulR)=\hat T_e(\dulr)+\hat V_{e,n}(\dulr,\dulR)$ is the BO Hamiltonian, containing the electronic kinetic energy $\hat T_e(\dulr)$ and all interactions $\hat V_{e,n}(\dulr,\dulR)$. As recently proven~\cite{AMG,AMG2}, the full wave function, $\Psi(\dulr,\dulR,t)$, solution of the TDSE
\begin{equation}\label{eqn: tdse}
 \hat H\Psi(\dulr,\dulR,t)=i\hbar\partial_t\Psi(\dulr,\dulR,t),
\end{equation}
can be written as the product 
\begin{equation}\label{eqn: factorization}
 \Psi(\dulr,\dulR,t)=\Phi_\dulR(\dulr,t)\chi(\dulR,t),
\end{equation}
of the nuclear wave function, $\chi(\dulR,t)$, and the electronic wave function, $\Phi_\dulR(\dulr,t)$, which parametrically depends of the nuclear configuration~\cite{hunter}. Throughout the paper the symbols $\dulr,\dulR$ indicate the coordinates of the $N_e$ electrons and $N_n$ nuclei, respectively. Eq.~(\ref{eqn: factorization}) is unique under the partial normalization condition (PNC)
\begin{equation}\label{eqn: pnc}
 \int d\dulr\left|\Phi_\dulR(\dulr,t)\right|^2=1\,\,\,\forall \,\dulR,t
\end{equation}
up to within a gauge-like phase transformation
\begin{equation}\label{eqn: gauge}
 \begin{array}{rcl}
  \chi(\dulR,t)\rightarrow\tilde\chi(\dulR,t)&=&e^{-\frac{i}{\hbar}\theta(\dulR,t)}\chi(\dulR,t) \\
  \Phi_\dulR(\dulr,t)\rightarrow\tilde\Phi_\dulR(\dulr,t)&=&e^{\frac{i}{\hbar}\theta(\dulR,t)}\Phi_\dulR(\dulr,t).
 \end{array}
\end{equation}
The evolution equations for $\Phi_\dulR(\dulr,t)$ and $\chi(\dulR,t)$,
\begin{align}
 \left(\hat H_{el}-\epsilon(\dulR,t)\right)\Phi_\dulR(\dulr,t) = i\hbar \partial_t\Phi_\dulR(\dulr,t)\label{eqn: electronic eqn} \\
 \hat H_n\chi(\dulR,t) = i\hbar \partial_t\chi(\dulR,t),\label{eqn: nuclear eqn}
\end{align}
are derived by applying Frenkel's action principle~\cite{kreibich1,frenkel,mclachlan} with respect to the two wave functions and are exactly equivalent~\cite{AMG,AMG2} to the TDSE~(\ref{eqn: tdse}). Eqs.~(\ref{eqn: electronic eqn}) and~(\ref{eqn: nuclear eqn}) are obtained by imposing the PNC~\cite{alonsoJCP2013,AMG2013} by means of Lagrange multipliers.

The electronic equation~(\ref{eqn: electronic eqn}) contains the electronic Hamiltonian
\begin{equation}
\hat H_{el}=\hat H_{BO}(\dulr,\dulR) + \hat U_{en}^{coup}[\Phi_\dulR,\chi],
\end{equation}
which is the sum of the BO Hamiltonian and the electron-nuclear coupling operator $\hat U_{en}^{coup}[\Phi_\dulR,\chi]$,
\begin{align}
\hat U_{en}^{coup}&[\Phi_\dulR,\chi]=\sum_{\nu=1}^{N_n}\frac{1}{M_\nu}\left[
 \frac{\left[-i\hbar\nabla_\nu-\bA_\nu(\dulR,t)\right]^2}{2} \right.\label{eqn: enco} \\
& \left.+\left(\frac{-i\hbar\nabla_\nu\chi}{\chi}+\bA_\nu(\dulR,t)\right)
 \left(-i\hbar\nabla_\nu-\bA_{\nu}(\dulR,t)\right)\right].\nonumber
\end{align}
In Eq.~(\ref{eqn: electronic eqn}), $\epsilon(\dulR,t)$ is the time-dependent PES (TDPES), defined as
\begin{equation}\label{eqn: tdpes I}
\epsilon(\dulR,t)=\left\langle \Phi_\dulR(t)\right|\hat H_{el}-i\hbar\partial_t\left|\Phi_\dulR(t)\right\rangle_\dulr.
\end{equation}
$\hat U_{en}^{coup}$ and $\epsilon(\dulR,t)$, along with the vector potential $\bA(\dulR,t)$,
\begin{equation}\label{eqn: vector potential}
\bA(\dulR,t) = \left\langle \Phi_\dulR(t)\right| \left. -i\hbar\nabla_\nu\Phi_\dulR(t)\right\rangle_\dulr,
\end{equation}
mediate the coupling between electrons and nuclei in a formally exact way. Here, the symbol
$\langle\,\cdot\,|\,\cdot\,\rangle_\dulr$ stands for an integration over electronic coordinates.

The nuclear evolution is generated by the Hamiltonian
\begin{equation}\label{eqn: nuclear H}
\hat H_n(\dulR,t) = \sum_{\nu=1}^{N_n} \frac{\left[-i\hbar\nabla_\nu+\bA_\nu(\dulR,t)\right]^2}{2M_\nu} + \epsilon(\dulR,t),
\end{equation}
according to the time-dependent Schr\"odinger-like equation~(\ref{eqn: nuclear eqn}). The nuclear wave function, by virtue of the PNC, reproduces the exact nuclear N-body density 
\begin{equation}
\Gamma(\dulR,t)=\left|\chi(\dulR,t)\right|^2=\int d\dulr\left|\Psi(\dulr,\dulR,t)\right|^2
\end{equation}
that is obtained from the full wave function. Moreover, the nuclear N-body current density can be directly obtained from $\chi(\dulR,t)$
\begin{equation} 
 {\bf J}_\nu(\dulR,t)=\frac{\Big[\mbox{Im}(\chi^*(\dulR,t)\nabla_\nu\chi(\dulR,t))+ \Gamma(\dulR,t){\bf A}_\nu(\dulR,t)\Big]}{M_\nu},
\end{equation}
thus allowing the interpretation of $\chi(\dulR,t)$ as a proper nuclear wave function.

The scalar and vector potentials are uniquely determined up to within the gauge transformations
\begin{equation}
 \begin{array}{rcl}
  \epsilon(\dulR,t)\rightarrow\tilde\epsilon(\dulR,t)&=&\epsilon(\dulR,t)+\partial_t\theta(\dulR,t) \\
  \bA(\dulR,t)\rightarrow\tilde \bA(\dulR,t)&=&\bA(\dulR,t)+\nabla_{\nu}\theta(\dulR,t).
 \end{array}
\end{equation}
The uniqueness can be straightforwardly proved by following the steps of the current density version~\cite{Ghosh-Dhara} of the  Runge-Gross theorem~\cite{RGT}. In this paper, as a choice of gauge, we introduce the additional constraint $\epsilon_{GD}(\dulR,t)=\langle\Phi_\dulR(t)|-i\hbar\partial_t\Phi_\dulR(t)\rangle_\dulr=0$, on the gauge-dependent component of the TDPES. Therefore, this scalar potential will be only expressed in terms of its gauge-invariant~\cite{long_steps, long_steps_mt} part, $\epsilon_{GI}(\dulR,t)$. It follows that the explicit expression of the TDPES is
\begin{align}
 \epsilon(\dulR,t)&=\epsilon_{GI}(\dulR,t)=\left\langle \Phi_{\dulR}(t)\right|\hat H_{BO}\left|\Phi_{\dulR}(t)\right\rangle_\dulr \label{eqn: tdpes}\\
 &+\sum_{\nu=1}^{N_n}\left[\frac{\hbar^2}{2M_\nu}\left\langle\nabla_\nu\Phi_{\dulR}(t)\right|\left.\nabla_\nu\Phi_{\dulR}(t)\right\rangle_\dulr-\frac{\bA_\nu^2(\dulR,t)}{2M_\nu}\right]\nonumber
\end{align}
where the second line is obtained from the action of the operator $\hat U_{en}^{coup}[\Phi_\dulR,\chi]$ in Eq.~(\ref{eqn: tdpes I}) on the electronic wave function. 

In the following, the electronic equation will be represented in the adiabatic basis. Therefore, it is worth introducing here the set of eigenstates $\lbrace\varphi_\dulR^{(j)}(\dulr)\rbrace$ of the BO Hamiltonian with eigenvalues $\epsilon_{BO}^{(j)}(\dulR)$. We expand the electronic wave function in this basis
\begin{equation}\label{eqn: BO expansion}
 \Phi_\dulR(\dulr,t) =\sum_jC_j(\dulR,t) \varphi_\dulR^{(j)}(\dulr)
\end{equation}
as well as the full wave function
\begin{equation}\label{eqn: BO expansion for Psi}
 \Psi(\dulr,\dulR,t) =\sum_jF_j(\dulR,t) \varphi_\dulR^{(j)}(\dulr).
\end{equation}
The coefficients of Eqs.~(\ref{eqn: BO expansion}) and (\ref{eqn: BO expansion for Psi}) are related by 
\begin{equation}\label{eqn: relation of the coefficients}
 C_j(\dulR,t)\chi(\dulR,t)=F_j(\dulR,t),
\end{equation}
which follows from Eq.~(\ref{eqn: factorization}). $|F_j(\dulR,t)|^2$ is interpreted as the amount of nuclear density that evolves ``on'' the $j$-th BO surface, as the nuclear density can be written as
\begin{equation}\label{eqn: chi as sum of F}
 \left|\chi(\dulR,t)\right|^2=\sum_j\left|F_j(\dulR,t)\right|^2,
\end{equation}
by using the PNC and the orthonormality of the adiabatic states. In this basis, the PNC reads
\begin{equation}\label{eqn: PNC on BO}
 \sum_j\left|C_j(\dulR,t)\right|^2=1\quad\forall\,\,\dulR,t.
\end{equation}

\section{The classical limit}\label{sec: limit}
The nuclear wave function, without loss of generality, can be written as~\cite{van-vleck}
\begin{equation}\label{eqn: van vleck}
\chi(\dulR,t)=\exp{\left[\frac{i}{\hbar}\mathcal S(\dulR,t)\right]},
\end{equation}
with $\mathcal S(\dulR,t)$ a complex function. We suppose now that this function can be expanded as an asymptotic series in powers of $\hbar$, namely $\mathcal S(\dulR,t)=\sum_{\alpha}\hbar^{\alpha}S_{\alpha}(\dulR,t)$. Inserting this expression in Eq.~(\ref{eqn: nuclear eqn}), the lowest order term, $S_0(\dulR,t)$, satisfies the equation
\begin{equation}\label{eqn: HJE}
 -\partial_t S_0(\dulR,t) = H_n\left(\dulR,\left\lbrace\nabla_\nu S_0(\dulR,t)\right\rbrace_{\nu=1,N_n},t\right),
\end{equation}
with $H_n$ defined as
\begin{equation}\label{eqn: classical hamiltonian}
 H_n = \sum_{\nu=1}^{N_n}\frac{\left[\nabla_\nu S_0(\dulR,t)+{\bf A}_\nu(\dulR,t)\right]^2}{2M_\nu}+ \epsilon(\dulR,t).
\end{equation}
Eq.~(\ref{eqn: HJE}) is obtained by considering only terms up to $\mathcal O(\hbar^0)$ and is formally identical to the Hamilton-Jacobi equation, if $S_0(\dulR,t)$ is identified with the classical action and, consequently, $\nabla_\nu S_0(\dulR,t)$ with the $\nu$th nuclear momentum,
\begin{equation}\label{eqn: classical momentum}
\nabla_\nu S_0(\dulR,t) = \bP_\nu.
\end{equation}
Therefore, $S_0(\dulR,t)$ is a real function and Eq.~(\ref{eqn: classical hamiltonian}) is the classical Hamiltonian corresponding to the quantum operator introduced in Eq.~(\ref{eqn: nuclear H}). It is worth noting that the canonical momentum derived from the classical Hamiltonian, as in the case of a classical charge moving in an electromagnetic field, is
\begin{equation}\label{eqn: canonical momentum}
 \widetilde{\bf P}_\nu(\dulR,t) = \nabla_\nu S_0(\dulR,t)+{\bf A}_\nu(\dulR,t).
\end{equation}
A more intuitive, less rigorous, step in the process of approximating the nuclei with classical particles is the identification of the nuclear density with a $\delta$-function~\cite{tully}, namely
\begin{equation}\label{eqn: classical density}
\left|\chi(\dulR,t)\right|^2 = \delta\left(\dulR-\dulR^{cl}(t)\right), 
\end{equation}
where the symbol $\dulR^{cl}(t)$ indicates the classical positions at time $t$. $\dulR^{cl}(t)$ is the classical trajectory. 

Eqs.~(\ref{eqn: classical momentum}) and~(\ref{eqn: classical density}) are the conditions for the nuclear degrees of freedom to behave classically. They can be obtained by performing the following limit operations
\begin{equation}\label{eqn: operations}
 \begin{array}{lc}
  \hbar\rightarrow 0 & \mathrm{(a)} \\
  \Sigma\rightarrow 0 & \mathrm{(b)}.
 \end{array}
\end{equation}
Eq.~(\ref{eqn: operations}a) follows from the fact that we consider only terms $\mathcal O(\hbar^0)$ in Eq.~(\ref{eqn: van vleck}). In Eq.~(\ref{eqn: operations}b), $\Sigma$ indicates the variance of a Gaussian-shaped nuclear density, centered at the classical positions $\dulR^{cl}(t)$, that becomes infinitely localized as in Eq.~(\ref{eqn: classical density}), when the limit operation is performed. The effect of Eqs.~(\ref{eqn: operations}a) and ~(\ref{eqn: operations}b) is
\begin{align}
\frac{-i\hbar\nabla_\nu\chi(\dulR,t)}{\chi(\dulR,t)}\rightarrow \nabla_\nu S_0(\dulR,t) & \,\mbox{ if }\, \hbar\rightarrow0 \label{eqn: evolution} \\
\left|C_j(\dulR,t)\right| \rightarrow \left|C_j(t)\right|& \,\mbox{ if }\, \Sigma\rightarrow0, \label{eqn: localization}
\end{align}
where the term $-i\hbar\nabla_{\nu}\chi/\chi$ appears explicitly in the definition of the electron-nuclear coupling operator given in Eq.~(\ref{eqn: enco}). In order to prove Eq.~(\ref{eqn: evolution}), we replace the nuclear wave function with its $\hbar$-expansion and we then take the limit $\hbar\rightarrow 0$. Only the zero-th order term survives, leading to Eq.~(\ref{eqn: evolution}), equivalent to
\begin{equation}
\frac{-i\hbar\nabla_\nu\chi(\dulR,t)}{\chi(\dulR,t)} = \bP_\nu.
\end{equation}
It will appear clear later that such term in the electronic equation is responsible for the non-adiabatic transitions induced by the coupling to the nuclear motion, as other MQC techniques, like the Ehrenfest method or the trajectory surface hopping~\cite{tully, barbatti, drukker}, also suggested. Here, we show that this term can be derived as the $\hbar\rightarrow 0$ limit in the exact equations, but it represents only the lowest order contribution in a $\hbar$-expansion. Moreover, this coupling is expressed via $\bP_\nu=\nabla_\nu S_0(\dulR,t)$, that is \textit{not} the canonical momentum appearing in the classical Hamiltonian (whose expression is given in Eq.~(\ref{eqn: canonical momentum})).

Eq.~(\ref{eqn: localization}) implies that the moduli of the coefficients $C_j(\dulR,t)$ in the expansion~(\ref{eqn: BO expansion}) become constant functions of $\dulR$ when the nuclear density is infinitely localized at the classical positions. Indeed, when the classical approximation strictly applies, the delocalization or the splitting of a nuclear wave packet is negligible ($\Sigma\rightarrow 0$). Therefore, any $\dulR$-dependence can be ignored and only the instantaneous classical position becomes relevant. It is worth underlining that this same hypothesis is at the basis of several MQC approaches~\cite{tully1990,tully,barbatti} and applies to the moduli and to the phases of the coefficients $C_j(\dulR,t)=|C_j(\dulR,t)|\exp{[(i/\hbar)\vartheta_j(\dulR,t)]}$. Here, we try to give a rigorous explanation of this assumption for the moduli, $\left|C_j(\dulR,t)\right|$, and we comment on the spatial dependence of the phases, $\vartheta_j(\dulR,t)$, in the following section.

Let us first suppose that $\chi(\dulR,t)$ is a normalized Gaussian wave packet, namely
\begin{equation}\label{eqn: nuclear wave packet}
 \left|\chi(\dulR,t)\right|^2 = G_{\Sigma}\left(\dulR-\dulR^{cl}(t)\right),
\end{equation}
centered at $\dulR^{cl}(t)$ with variance $\Sigma$. In the classical limit, $|\chi(\dulR,t)|^2$ reduces to a $\delta$-function at $\dulR^{cl}(t)$, consequently at each point $\dulR$ where $|\chi(\dulR,t)|^2$ is zero, all terms on the right-hand-side of Eq.~(\ref{eqn: chi as sum of F}) have to be zero, since they are all non-negative. Therefore, $|F_j(\dulR,t)|^2$ should become $\delta$-functions at $\dulR^{cl}(t)$ $\forall \,j$. Since we are interested in this limit, we represent each $|F_j(\dulR,t)|^2$ by a not-normalized Gaussian ($F_j(\dulR,t)$ is not normalized), centered at different positions, $\dulR^j(t)$, than $\dulR^{cl}(t)$. Using this hypothesis, Eq.~(\ref{eqn: chi as sum of F}) becomes 
\begin{equation}\label{eqn: equality of Gs}
 G_{\Sigma}\left(\dulR-\dulR^{cl}(t)\right)=\sum_jB_j^2(t)G_{\sigma_j}\left(\dulR-\dulR^j(t)\right),
\end{equation}
where $\sum_j B_j^2(t)=1$ accounts for the normalization of $\chi(\dulR,t)$. The pre-factors $B_j^2(t)$ have been introduced because each $F_j(\dulR,t)$ is not normalized, $G_{\sigma_j}(\dulR-\dulR^j(t))$ is instead a normalized Gaussian centered at $\dulR^j(t)$ with variance $\sigma_j$. The variances $\Sigma$ and $\sigma_j$ are allowed to be time-dependent even if we are not explicitly indicating this dependence. We will prove the following statements
\begin{align}
\mbox{(i)} &\,\, \Sigma=\sigma_j \,\,\forall j,\,t \label{eqn: statement i}\\
\mbox{(ii)}&\,\, \dulR^{cl}(t)=\dulR^j(t) \,\,\forall j,\,t. \label{eqn: statement ii}
\end{align}
To this end, we compare the behavior of both sides of Eq.~(\ref{eqn: equality of Gs}) for $\dulR\rightarrow \pm\infty$: we need to show that
\begin{equation}\label{eqn: limit R inf}
 1 = \lim_{\dulR\rightarrow\pm \infty}\sum_j B_j^2(t)\frac{\Sigma}{\sigma_j}
e^{\left[-\frac{\left(\dulR-\dulR^j(t)\right)^2}{\sigma_j^2}+\frac{\left(\dulR-\dulR^{cl}(t)\right)^2}{\Sigma^2}\right]},
\end{equation}
where we wrote explicitly the expressions of the Gaussian functions in Eq.~(\ref{eqn: equality of Gs}). The leading term on the right-hand-side of Eq.~(\ref{eqn: limit R inf}) has to satisfy the condition 
\begin{equation}\label{eqn: finite limit in R}
 \lim_{\left|\dulR\right|\rightarrow \infty}
 e^{-\left(\sigma_j^{-2}-\Sigma^{-2}\right)\left|\dulR\right|^2}< \infty\quad\forall j
\end{equation}
or, equivalently,
\begin{equation}\label{eqn: relation for the variance 1}
 \sigma_j^{-2}-\Sigma^{-2} \geq 0 \Rightarrow \sigma_j^2 \leq \Sigma^2\quad\forall j.
\end{equation}
A similar argument is applied to the Fourier Transform ($FT$) of both sides of Eq.~(\ref{eqn: equality of Gs})
\begin{equation}
 \hat G_{\tilde{\Sigma},\dulR^{cl}}(\dulk) = \sum_j B_j^2(t)\hat G_{\tilde{\sigma}_j,\dulR_j}(\dulk),
\end{equation}
where
\begin{equation}
 \hat G_{\tilde{\Sigma},\dulR^{cl}}(\dulk)=FT\left[G_{\Sigma}(\dulR-\dulR^{cl})\right](\dulk)=\frac{e^{i\dulR^{cl}\cdot\dulk}}{2\pi} e^{-\tilde{\Sigma}^2\left|\dulk\right|^2}
\end{equation}
(and similarly for $\hat G_{\tilde{\sigma}_j,\dulR^j}(\dulk)$) with $\tilde{\Sigma}=\Sigma/2$. The Gaussian transforms to another Gaussian with inverse variance, and if we calculate the limit $|\dulk|\rightarrow\infty$, we obtain a relation similar to Eq.~(\ref{eqn: relation
for the variance 1}), namely
\begin{equation}\label{eqn: relation for the variance 2}
 \tilde{\sigma}_j^2-\tilde{\Sigma}^2 \geq 0 \Rightarrow \sigma_j^2 \geq \Sigma^2\quad\forall j.
\end{equation}
Eqs.~(\ref{eqn: relation for the variance 1}) and (\ref{eqn: relation for the variance 2}) are simultaneously satisfied if
\begin{equation}\label{eqn: equality for the variances}
 \sigma_j^2=\Sigma^2 \Rightarrow \sigma_j=\Sigma\quad\forall j,
\end{equation}
and this proves statement~(\ref{eqn: statement i}). In order to prove statement~(\ref{eqn: statement ii}), we study the behavior of Eq.~(\ref{eqn: equality of Gs}) at $\dulR^{cl}(t)$, namely
\begin{equation}
 1= \sum_j B_j^2(t) e^{-\frac{\left(\dulR^{cl}(t)-\dulR^j(t)\right)^2}{\Sigma^2}}.
\end{equation}
Since the pre-factors $B_j^2(t)$ sum up to one, the relation
\begin{equation}\label{eqn: equality for the mean value}
 \dulR^j(t)=\dulR^{cl}(t)
\end{equation}
must hold, since $0\leq \exp{[-[\dulR^{cl}(t)-\dulR^j(t)]^2/\Sigma^2]} < 1$ if $\dulR^j(t)\neq\dulR^{cl}(t)$. Using Eq.~(\ref{eqn: relation of the coefficients}), we can now show that
\begin{equation}\label{eqn: modulus of cj constant in space}
 \left|C_j(\dulR,t)\right| =
 \left[\frac{B_j^2(t)G_{\sigma_j}\left(\dulR-\dulR^j(t)\right)}{G_{\Sigma}\left(\dulR-\dulR^{cl}(t)\right)}\right]^{\frac{1}{2}} = B_j(t),
\end{equation}
or in other words $|C_j(\dulR,t)|$ is only a function of time and is constant in space. It is worth noting that Eqs.~(\ref{eqn: equality for the variances}) and (\ref{eqn: equality for the mean value}) have to be valid at all times. 

As a consequence of the discussion presented so far, we obtain that the exact (quantum mechanical) population of the BO states as function of time
\begin{equation}\label{eqn: exact j-th population}
\rho_j(t) = \int d\dulR \left|F_j(\dulR,t)\right|^2 
\end{equation}
must equal the population calculated as
\begin{equation}\label{eqn: mqc j-th population}
 \left|C_j(t)\right|^2 = \int d\dulR\left|C_j(t)\right|^2\delta\left(\dulR-\dulR^{cl}(t)\right),
\end{equation}
when the classical limit is performed.

\subsection{Spatial dependence of the phases}\label{sec: phases}
We will discuss in this section why the phases $\vartheta_j(\dulR,t)$ of the coefficients $C_j(\dulR,t)$ in Eq.~(\ref{eqn: BO expansion}) will be considered constant functions of $\dulR$, similarly to the moduli $|C_j(\dulR,t)|$ as shown in the previous section. We mention again that this hypothesis is usually introduced in the derivation of other MQC approaches~\cite{tully1990,tully,barbatti} and we try here to justify this approximation. 


First of all, we introduce the expression of the vector potential in the adiabatic basis
\begin{align}
 \bA_\nu(\dulR,t)=&-i\hbar\sum_{j,k}C_j^*(\dulR,t)C_k(\dulR,t) \left\langle
 \varphi_\dulR^{(j)}\right|\left.\nabla_{\nu}\varphi_\dulR^{(k)}\right\rangle_\dulr \nonumber \\
 &+\sum_j\left|C_j(\dulR,t)\right|^2\nabla_{\nu}\vartheta_j(\dulR,t)
 \label{eqn: vector potential on the adiabatic basis}
\end{align}
and we define the quantities
\begin{eqnarray}
 \bA_\nu'(\dulR,t)&=&-i\hbar\sum_{j,k}C_j^*(\dulR,t)C_k(\dulR,t){\bf d}_{jk,\nu}^{(1)}(\dulR)\label{eqn: A'} \\
 \bA_\nu''(\dulR,t)&=&\sum_j\left|C_j(\dulR,t)\right|^2\nabla_{\nu}\vartheta_j(\dulR,t).\label{eqn: A''} 
\end{eqnarray}
We used here the definition of the first order non-adiabatic coupling (NAC) ${\bf d}_{jk,\nu}^{(1)}(\dulR)=
\langle \varphi_\dulR^{(j)}|\nabla_\nu\varphi_\dulR^{(k)}\rangle_\dulr$. Moreover, the symbol $\phi_j(\dulR,t)$ will be used to indicate the phase of the coefficient $F_j(\dulR,t)$, of the expansion~(\ref{eqn: BO expansion for Psi}).

We will show now that the dependence on the index $j$ of the phases $\vartheta_j(\dulR,t)$ can be dropped. This is done by relating the quantity $\nabla_\nu\vartheta_j(\dulR,t)$ to the rate of displacement of the mean value(s) of $|\chi(\dulR,t)|$ and $|F_j(\dulR,t)|$. We are going to use again the hypothesis of a Gaussian-shaped nuclear density, which infinitely localizes at $\dulR^{cl}(t)$ in the classical limit.

Two equivalent exact expressions for the expectation value of the nuclear momentum are obtained from Eqs.~(\ref{eqn: factorization}) and (\ref{eqn: BO expansion for Psi}), namely
\begin{align}
&\int d\dulR \left[\nabla_\nu S(\dulR,t)+\bA_\nu(\dulR,t)\right]\left|\chi(\dulR,t)\right|^2 = \\
&\int d\dulR \left[\sum_j\left|C_j(\dulR,t)\right|^2\nabla_{\nu}\phi_j(\dulR,t)+\bA_\nu'(\dulR,t)\right]\left|\chi(\dulR,t)\right|^2,\nonumber
\end{align}
where $S(\dulR,t)$ is the phase of $\chi(\dulR,t)$. The identity of the terms in square brackets under the integral signs follows,
\begin{equation}\label{eqn: relation between nuclear momenta}
 \dulP_\nu(\dulR,t)+\bA_\nu(\dulR,t) = \sum_j\left|C_j(t)\right|^2\nabla_{\nu}\phi_j(\dulR,t)+\bA_\nu'(\dulR,t),
\end{equation}
where the approximations used here are (i) replacing $S(\dulR,t)$ with $S_0(\dulR,t)$ from Eq.~(\ref{eqn: HJE}) (then using Eq.~(\ref{eqn: classical momentum}) for the nuclear momentum) and (ii) neglecting the spatial dependence of $|C_j(\dulR,t)|$ (as proven, in the previous section, to be consistent with the classical treatment of the nuclei).  As will be derived in Appendix~\ref{app: classical H}, the term on the left-hand-side of Eq.~(\ref{eqn: relation between nuclear momenta}) can be also written as
\begin{equation}\label{eqn: use of the real momentum}
 \dulP_\nu(\dulR,t)+\bA_\nu(\dulR,t)=M_\nu\dot{\bf R}^{cl}_\nu(t)
\end{equation}
where $\dot{\bf R}^{cl}_\nu(t)$ $\forall \, \nu$ is the displacement rate of $\dulR^{cl}(t)$, the mean value of the nuclear density. If we define the quantity $\dulP_\nu^{j}(\dulR,t)=\nabla_{\nu}\phi_j(\dulR,t)$, Eq.~(\ref{eqn: relation between nuclear momenta}) can be re-written as 
\begin{equation}\label{eqn: term in parenthesis}
 M_\nu\dot{\bf R}^{cl}_\nu(t)=\sum_j\left|C_j(t)\right|^2\Big(\dulP_\nu^{j}(\dulR,t)+\bA_\nu'(\dulR,t)\Big),
\end{equation}
where the PNC $\sum_j|C_j(t)|^2=1$ has been used. By analogy with Eq.~(\ref{eqn: use of the real momentum}), the term in parenthesis can be defined as
\begin{equation}
M_\nu\dot{\bf R}^j_\nu(t) = \dulP_\nu^{j}(\dulR,t)+\bA_\nu'(\dulR,t), 
\end{equation}
the momentum associated to the motion of the mean value $\dulR^j(t)$ ($\forall\,\nu$) of the BO-projected wave packet $F_j(\dulR,t)$. This is also consistent with our previous statement~\cite{long_steps_mt} on the connection between the phase $\phi_j(\dulR,t)$ and the propagation velocity of $\dulR^j(t)$ based on semi-classical arguments. Therefore, we obtain
\begin{equation}\label{eqn: equality for the velocities}
 \dot\dulR^{cl}(t)=\sum_j\left|C_j(t)\right|^2 \dot\dulR^j(t),
\end{equation}
but since Eq.~(\ref{eqn: equality for the mean value}) states that $\dulR^{cl}(t)=\dulR^j(t)$ $\forall \,t$, the equality $\dot\dulR^{cl}(t)=\dot\dulR^j(t)$ holds. Eq.~(\ref{eqn: equality for the velocities}) becomes an identity due to the PNC. 

We have shown that the term in parenthesis in Eq.~(\ref{eqn: term in parenthesis}) is independent of $j$, namely
\begin{equation}\label{eqn: nabla phi_j}
 \nabla_{\nu}\phi_j(\dulR,t)=\nabla_{\nu}\phi(\dulR,t) \quad \forall\,j.
\end{equation}
If we also use Eq.~(\ref{eqn: relation of the coefficients}), then $\forall \,j$ the relation
\begin{align}
 \nabla_{\nu}\vartheta_j(\dulR,t)&=\nabla_{\nu} S_0(\dulR,t)-\nabla_{\nu}\phi_j(\dulR,t) \nonumber \\
 &=\nabla_{\nu} S_0(\dulR,t)-\nabla_{\nu}\phi(\dulR,t)=\nabla_{\nu}\vartheta(\dulR,t)\label{eqn: theta indpendent of j}
\end{align}
holds. This result is used in the gauge condition, which we recall here
\begin{equation}
 \epsilon_{GD}(\dulR,t)=\left\langle\Phi_\dulR(t)\right|-i\hbar\partial_t\left|\Phi_\dulR(t)\right\rangle_\dulr=0.
\end{equation}
In the adiabatic basis, it becomes
\begin{equation}\label{eqn: gauge in BO basis}
 \sum_j \left|C_j(t)\right|^2\partial_t\vartheta_j(\dulR,t)=0.
\end{equation}
Within the classical treatment of nuclear dynamics, the chain rule~\cite{tully1990} $\partial_t=\sum_\nu\dot{\bf R}_\nu\cdot\nabla_{\nu}$ can be used, due to the relation between time and nuclear space represented by the trajectory $\dulR,t\rightarrow\dulR^{cl}(t)$. Therefore, the gauge condition~(\ref{eqn: gauge in BO basis}), together with Eq.~(\ref{eqn: theta indpendent of j}), leads to the relation
\begin{equation}\label{eqn: gauge and dR theta}
\sum_\nu \dot{\bf R}_\nu \cdot \nabla_{\nu}\vartheta(\dulR,t)=0.
\end{equation}
If $\dot{\bf R}_\nu\neq0$, from Eq.~(\ref{eqn: gauge and dR theta}) we derive
\begin{equation}\label{eqn: constant phases}
\nabla_\nu\vartheta\left(\dulR,t\right) =0
\end{equation}
since no particular relation among $\nabla_{\nu}\vartheta(\dulR,t)$ for different values of the index $\nu$ exists, that guarantees that the sum is exactly zero. At the classical turning points, where the nuclear velocity is zero, the value of $\nabla_{\nu}\vartheta(\dulR,t)$ cannot be defined by Eq.~(\ref{eqn: gauge and dR theta}). In this case, we will anyway consider Eq.~(\ref{eqn: constant phases}) valid.

The main results obtained from this discussion are
\begin{equation}\label{eqn: localization bis}
 C_j(\dulR,t)\rightarrow C_j(t),
\end{equation}
to be used in the electronic evolution equation~(\ref{eqn: electronic eqn}) when the wave function $\Phi_\dulR(\dulr,t)$ is expanded on the adiabatic basis, and the approximated expression of the vector potential
\begin{equation}\label{eqn: final expression vector potential with d}
 \bA_\nu(\dulR,t)=-i\hbar\sum_{j,k}C_j^*(t)C_k(t){\bf d}_{jk,\nu}^{(1)}(\dulR),
\end{equation}
consistent with the classical limit (since $\bA_\nu''(\dulR,t)$ in Eq.~(\ref{eqn: A''}) is identically zero).

Eq.~(\ref{eqn: localization bis}), however, breaks the invariance of the electron-nuclear dynamics under a gauge transformation. For instance, the force term $\nabla_{\nu}\epsilon_{GD}(\dulR,t)=0$ independently of the choice of the gauge, where we recall that $\epsilon_{GD}(\dulR,t)=\langle\Phi_{\dulR}(t)\vert-i\hbar\partial_t\Phi_{\dulR}(t)\rangle_\dulr$ is the gauge-dependent component~\cite{long_steps} of the TDPES in Eq.~(\ref{eqn: tdpes}). It follows that a gauge, compatible with the approximation $C_j(\dulR,t)=C_j(t)$, has to be chosen, like the gauge adopted in our calculations $\epsilon_{GD}(\dulR,t)=0$.

\section{Mixed quantum-classical equations}\label{sec: mqc}
The effect of performing the classical limit in Eqs.~(\ref{eqn: electronic eqn}) and~(\ref{eqn: nuclear eqn}) will be used here to determine the MQC equations of motion as the lowest order approximation to the exact decomposition of electronic and nuclear motion in the framework of the factorization of the full wave function.

The classical trajectory is determined by Newton's equation~\cite{chin}
\begin{align}
 \dot{\widetilde{\bf P}}_\nu =- \nabla_{\nu}\epsilon+ \partial_t{\bf A}_\nu-{\bf V}_{\nu}\times{\bf B}_{\nu\nu}+\sum_{\nu'\neq\nu} {\bf F}_{\nu\nu'},\label{eqn: classical evolution}
\end{align}
with ${\bf V}_{\nu}=\widetilde{\bf P}_\nu/M_\nu$. As in~\cite{mqc}, Eq.~(\ref{eqn: classical evolution}) is derived by acting with the gradient operator $\nabla_{\nu}$ on Eq.~(\ref{eqn: HJE}) and by identifying the total time derivative operator as $\partial_t+\sum_{\nu'}{\bf V_{\nu'}\cdot \nabla_{\nu'}}$.It can be easily proven that Eq.~(\ref{eqn: classical evolution}) is invariant under a gauge transformation: also in the classical approximation, the force produced by the vector and scalar potentials maintains its gauge-invariant property, as indeed happens in the exact quantum treatment. Henceforth, all quantities depending on $\dulR,t$ become functions of $\dulR^{cl}(t)$, the classical path along which the action $S_0(\dulR^{cl}(t))$ is stationary.

The first three terms on the right-hand-side produce the electromagnetic force due to the presence of the vector and scalar potentials, with ``generalized'' magnetic field
\begin{equation}\label{eqn: generalized magnetic field}
{\bf B}_{\nu\nu'}(\dulR^{cl}(t))=\nabla_{\nu}\times {\bf A}_{\nu'}(\dulR^{cl}(t)).
\end{equation}
The remaining term
\begin{align}
 {\bf F}_{\nu\nu'}(\dulR^{cl}(t)) &= -{\bf V}_{\nu'}\times{\bf B}_{\nu\nu'}(\dulR^{cl}(t)) \label{eqn: off-diagonal force}\\
  +&\left[\left({\bf V}_{\nu'}\cdot\nabla_{\nu'}\right){\bf A}_{\nu}(\dulR^{cl}(t))- \left({\bf V}_{\nu'}\cdot\nabla_{\nu}\right){\bf
 A}_{\nu'}(\dulR^{cl}(t))\right]\nonumber
\end{align}
is an inter-nuclear force term, arising from the coupling with the electronic system. Eq.~(\ref{eqn: off-diagonal force}) shows the non-trivial effect of the vector potential on the classical nuclei~\cite{lu1,lu2}, as it does not only appear in the bare electromagnetic force, but also ``dresses'' the nuclear interactions. In those cases where the vector potential is curl-free, the gauge can be chosen by setting the vector potential to zero, then Eqs.~(\ref{eqn: generalized magnetic field}) and~(\ref{eqn: off-diagonal force}) are identically zero. Only the component of the vector potential that is not curl-free cannot be gauged away. Whether and under which conditions $\mbox{curl}\,\bA_\nu(\dulR,t)=0$ is, at the moment, subject of investigations~\cite{CI_MAG}.

The coupled partial differential equations for the coefficients $C_j(\dulR,t)$ in the expansion of the electronic wave function on the adiabatic states simplify to a set of ordinary differential equations in the time variable only
\begin{equation}\label{eqn: el ode}
\dot C_j(t)=-\frac{i}{\hbar}[\epsilon^{(j)}_{BO}-\epsilon]C_j(t)+\sum_k C_k(t) U_{jk},
\end{equation}
where all quantities depending on $\dulR$, as $\epsilon^{(j)}_{BO}$, $\epsilon$ and $U_{jk}$, have to be evaluated at the instantaneous nuclear position. The symbol $U_{jk}$ is used to indicate the matrix elements (times $-i/\hbar$) of the operator $\hat U_{en}^{coup}[\Phi_\dulR,\chi]$ on the adiabatic basis. Its expression, using the first and second order $d_{jk,\nu}^{(2)}(\dulR)=\langle\nabla_{\nu}\varphi_\dulR^{(j)}|\nabla_{\nu}\varphi_\dulR^{(k)}\rangle_\dulr$ NACs, is
\begin{align}
U_{jk}&=\sum_\nu\frac{\delta_{jk}}{M_\nu}\left[\frac{i}{\hbar}\left(\frac{{\bf A}_{\nu}^2}{2}+{\bf A}_\nu\cdot \nabla_\nu S_0\right)+\frac{\nabla_{\nu}\cdot {\bf A}_\nu}{2}\right] \nonumber \\
&-\sum_\nu\frac{1}{M_\nu}\left[{\bf d}_{jk,\nu}^{(1)}\cdot\nabla_\nu S_0-\frac{i\hbar}{2}\left(\nabla_{\nu}\cdot{\bf d}_{jk,\nu}^{(1)}-d_{jk,\nu}^{(2)}\right)\right].\label{eqn: U on BO states}
\end{align}
Similarly, the TDPES can be expressed on the adiabatic basis as
\begin{align}
&\epsilon(\dulR,t)=\sum_j\left|C_j(t)\right|^2\epsilon_{BO}^{(j)}+i\hbar\sum_{j,k}C_j^*(t)C_k(t)U_{jk}
\label{eqn: tdpes on BO states}
\end{align}
while the vector potential is given by Eq.~(\ref{eqn: final expression vector potential with d}).

The electronic evolution equation~(\ref{eqn: el ode}) contains three different contributions: (i) a diagonal oscillatory term, given by the expression in square brackets in Eq.~(\ref{eqn: el ode}) plus the term in parenthesis in the first line of Eq.~(\ref{eqn: U on BO states}); (ii) a diagonal sink/source term, arising from the divergence of the vector potential in Eq.~(\ref{eqn: U on BO states}), that may cause exchange of populations between the adiabatic states even if off-diagonal couplings are neglected; (iii) a non-diagonal term inducing transitions between BO states, that contains a dynamical term proportional to the nuclear momentum (first term in the second line of Eq.~(\ref{eqn: U on BO states})), as suggested in other QC approaches~\cite{tully,barbatti,drukker} and a term containing the second order NACs. In particular, the dynamical non-adiabatic contribution follows from the classical approximation in Eq.~(\ref{eqn: evolution}) and drives the electronic population exchange induced by the motion of the nuclei.

The MQC scheme derived here introduces new contributions, both in the electronic and in the nuclear equations of motion, if compared to the Ehrenfest approach (see for instance the first line in Eq.~(\ref{eqn: U on BO states}) or the second term on the right-hand-side of Eq.~(\ref{eqn: tdpes on BO states})). The study of the actual extent and the effect of this corrections to Ehrenfest dynamics is beyond the scope of this paper and shall be addressed by investigating a wider class of problems than the simple model presented here. For instance, it would be interesting to analyse those situations where the vector potential plays an important role. This analysis goes hand-in-hand with our ongoing investigation~\cite{CI_MAG} for cases where this exact vector potential cannot be gauged away.

\section{Numerical results}\label{sec: results}
\begin{figure}
 \begin{center}
  \includegraphics*[width=.6\textwidth]{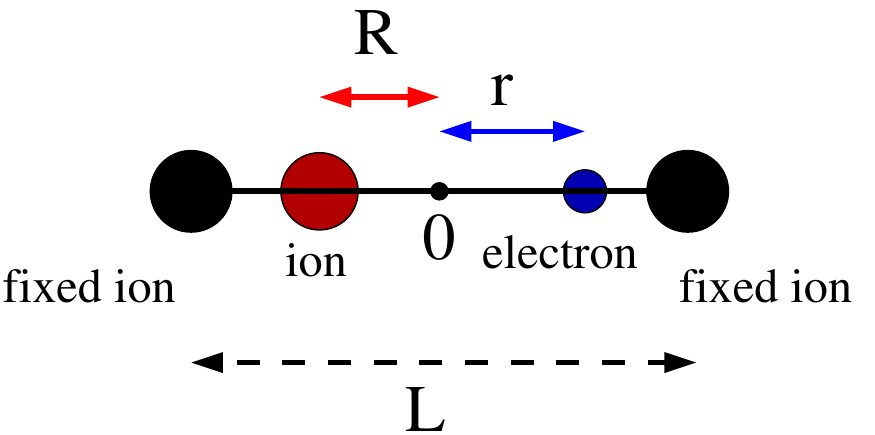}
 \end{center}
 \caption{Schematic representation of the 1D model system whose Hamiltonian is given in Eq.~(\ref{eqn: metiu}). The red ion moves between the fixed ions and the electron, which interacts with all the ions via a soft Coulomb potential, is allowed to move beyond the fixed ions. $R$ is the ion position and $r$ the electron position with respect to the origin 0. $L$ is the distance between the fixed ions.}
 \label{fig: system} 
\end{figure}

\begin{figure}
 \begin{center}
  \includegraphics*[angle=270,width=.7\textwidth]{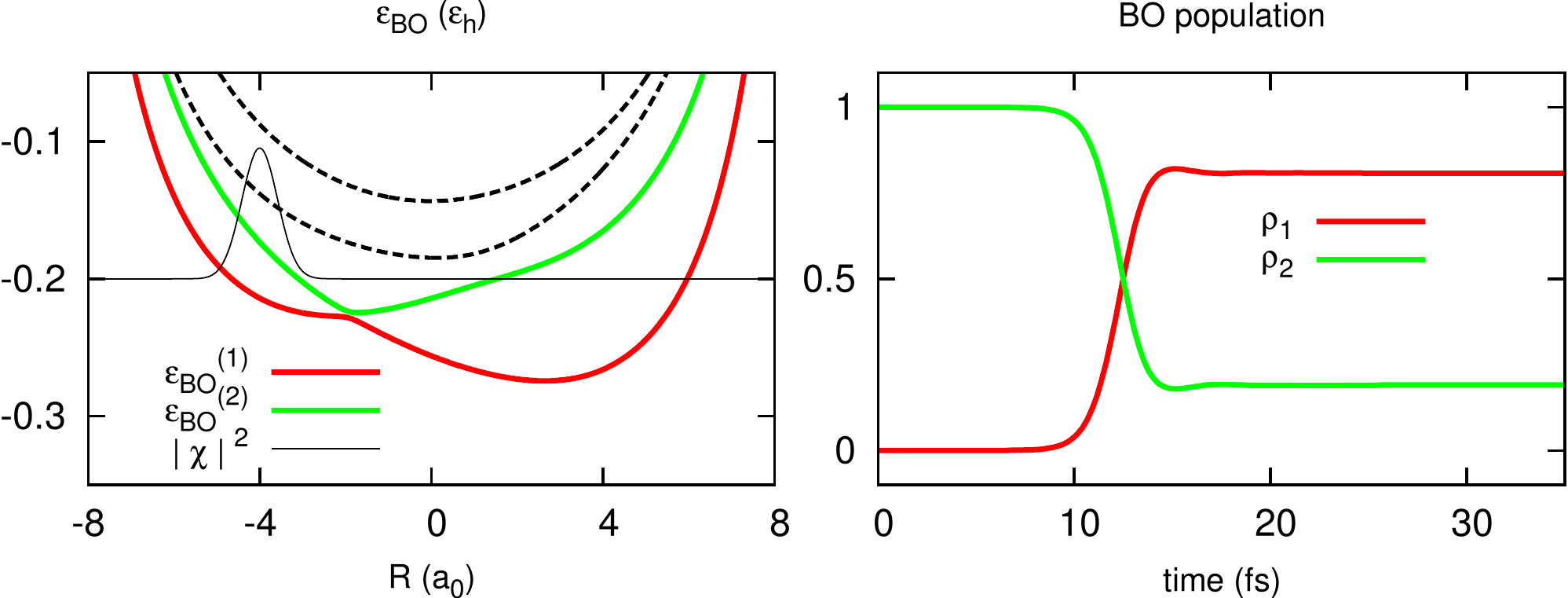}
 \end{center}
 \caption{Left: first (red line) and second (green line) BO surfaces and initial Gaussian wave packet (thin black line) centered at $R_0$. The third and fourth BO surfaces (dashed black lines) are shown for reference. Right: populations of the BO states $\rho_1$ (red line) and $\rho_2$ (green line) as functions of time, from exact calculations.}
 \label{fig: model} 
\end{figure}

We employ this new MQC scheme to study a model that is simple enough to allow for an exact treatment, by solving the TDSE, but at the same time exhibits characteristic features associated with non-adiabatic dynamics. It was originally developed by Shin and Metiu~\cite{metiu} to study charge transfer processes and consists of three ions and a single electron. Two ions are fixed at a distance $L=19.0$~a$_0$, the third ion and the electron are free to move in one dimension along the line joining the two fixed ions. A schematic representation of the system is shown in Fig.~\ref{fig: system}.
The Hamiltonian of  this system reads
\begin{align}
 \hat{H}(r,R) =  - \frac{1}{2}\frac{\partial^2}{\partial r^2}-\frac{1}{2M}\frac{\partial^2}{\partial R^2}  + \frac{1}{\vert \frac{L}{2} -R \vert } + \frac{1}{\vert \frac{L}{2} + R \vert} \nonumber\\
 - \frac{\mathrm{erf}\left(\frac{\vert R-r \vert}{R_f}\right)}{\vert R-r \vert} - \frac{\mathrm{erf}\left(\frac{\vert r-\frac{L}{2}
 \vert}{R_r}\right)}{\vert r-\frac{L}{2}  \vert} -
 \frac{\mathrm{erf}\left(\frac{\vert r+\frac{L}{2} \vert}{R_l}\right)}{\vert r+\frac{L}{2} \vert},\label{eqn: metiu}
\end{align}
where the symbols $r,R$ have been used for the positions of the electron and the ion in one dimension. 
Here, $M=1836$, the proton mass, and $R_f=5.0$~a$_0$, $R_l=3.1$~a$_0$ and $R_r=4.0$~a$_0$, such that the first adiabatic potential energy surface, $\epsilon_{BO}^{(1)}$, is coupled to the second, $\epsilon_{BO}^{(2)}$, and the two are decoupled from the rest of the surfaces, i.e. the dynamics of the system can be described by considering only two adiabatic states. The BO surfaces are shown in Fig.~\ref{fig: model} (left), where energies are expressed in Hartree ($\epsilon_h$). Henceforth, we will drop the bold-double underlined notation for electronic and nuclear positions as we are dealing with one dimensional quantities.

The initial condition for quantum propagation is $\Psi(r,R,0) = G_{\Sigma}^{1/2}(R-R_0)\varphi_R^{(2)}(r)$, where $G_{\Sigma}^{1/2}(R-R_0)$ is a real Gaussian ($G_{\Sigma}$ is normalized to unity) with variance $\Sigma=1/\sqrt{2.85}$~$a_0$ centered at $R_0=-4.0$~$a_0$ and $\varphi_R^{(2)}(r)$ is the first excited BO state. The TDSE is solved, numerically, using the split operator technique~\cite{spo}, with time step $2.4\times10^{-3}$~fs ($0.1$~a.u.).

\subsection{Validity of the classical approximation}\label{sec: results 1}
The classical approximation of nuclear dynamics is strictly valid if the nuclear density remains
localized at the classical position $R^{cl}(t)$. Due to the fact that $|\chi(R,t)|^2$ is the sum of contributions, or
partial densities, propagating ``on'' different BO surfaces, the localization condition should also apply to each
contribution, as discussed in Section~\ref{sec: limit}. Moreover, we observed in Section~\ref{sec: phases} that the displacement of the nuclear wave packet, expressed as the displacement of its mean position, shall be the same as the displacements of the mean positions associated to $|F_j(R,t)|^2$. We calculate and compare these quantities, in order to predict agreement/deviation of the results from MQC calculation (shown in Section~\ref{sec: results 2}) with/from the results from quantum calculations.

We compare the mean nuclear position as function of time, given by the expression 
\begin{equation}\label{eqn: R_n}
 R_n(t) = \int dR R \left|\chi(R,t)\right|^2,
\end{equation}
with the mean positions calculated with $|F_j(R,t)|^2$
\begin{equation}
 R^{\mathrm{qm}}_j(t) = \frac{1}{\rho_j(t)}\int dR R \left|F_j(R,t)\right|^2
\end{equation}
with the normalization factor $\rho_j(t)$ from Eq.~(\ref{eqn: exact j-th population}). As long as $R_n(t)$ and $R^{\mathrm{qm}}_j(t)$ are close to each other, agreement between the exact and approximated propagation schemes is expected.

\begin{figure}[h!]
 \begin{center}
  \includegraphics*[angle=270,width=.6\textwidth]{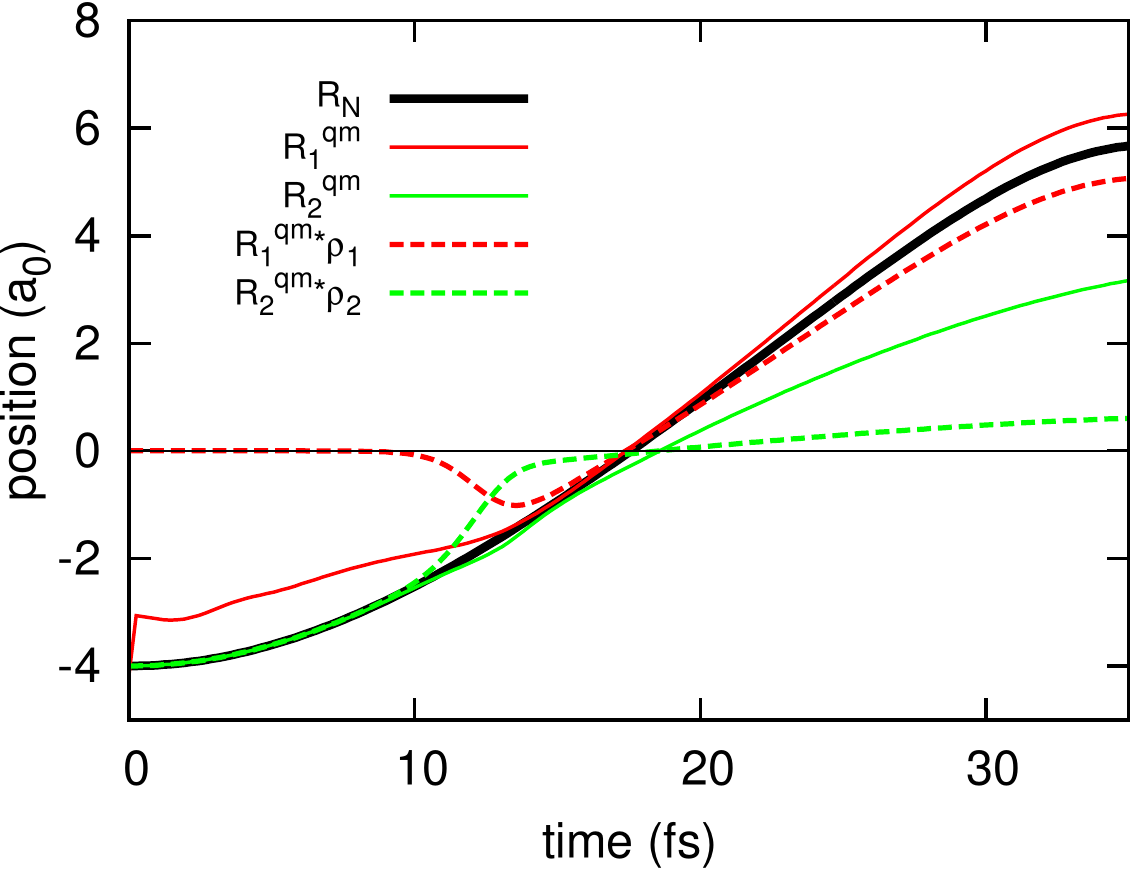}
 \end{center}
 \caption{$R_n$ (black line), $R^{\mathrm{qm}}_j(t)$ ($j=1$ red line, $j=2$ green line) and $R^{\mathrm{qm}}_j(t)\rho_j(t)$ ($j=1$ dashed red line, $j=2$ dashed green line) as functions of time.  The thin horizontal line shows the value 0.}
 \label{fig: positions} 
\end{figure}

Fig.~\ref{fig: positions} shows, as a thick black line, the mean nuclear position $R_n(t)$. The continuous red and green lines, respectively the mean positions of the wave packets propagating on $\epsilon_{BO}^{(1)}(R)$ and on $\epsilon_{BO}^{(2)}(R)$, considerably deviate from $R_n(t)$ only after 20~fs. This suggests a deviation of the nuclear evolution from a purely classical behavior. However, if we now compare the thick black line with the dashed red and green lines, we observe a different behavior: the dashed green line coincides with the black line up to 10~fs and after 15~fs the dashed red line coincides and remains close to the black line until the end of the simulated dynamics. 
The dashed lines represent the positions $R^{\mathrm{qm}}_j(t)$ weighted by the corresponding populations $\rho_j(t)$ of the BO states. Indeed, the expression of the mean nuclear position in Eq.~(\ref{eqn: R_n}) can be written as
\begin{equation}
 R_n(t) = \rho_1(t) R^{\mathrm{qm}}_1(t)+\rho_2(t) R^{\mathrm{qm}}_2(t),
\end{equation}
where Eq.~(\ref{eqn: chi as sum of F}) has been used, and in Fig.~(\ref{fig: positions}) we observe the following relations
\begin{equation}\label{eqn: mean nuclear position}
 R_n(t) \simeq \left\lbrace
 \begin{array}{cc}
 \rho_2(t) R^{\mathrm{qm}}_2(t), & t<10\textrm{~fs} \\
 & \\
 \rho_1(t) R^{\mathrm{qm}}_1(t), & t>15\textrm{~fs}.
 \end{array}\right.
\end{equation}
This is an expected results, due to strong non-adiabatic nature of the process, as shown in Fig.~\ref{fig: model} (right). This property may suggest a good agreement between exact and MQC results if we compare the mean values extracted from the two propagation schemes. As we will show below, when a single trajectory scheme is used to evolve classical nuclei according the MQC scheme proposed here, the trajectory is able to visit those regions of space with the largest probability of finding the (quantum) particle. This is what we have presented in Ref.~\cite{long_steps}, by evolving a single trajectory on the exact TDPES. In the example discussed here, it will be shown that the classical particle, propagating according to the force in Eq.~(\ref{eqn: classical evolution}), tracks $R^{\mathrm{qm}}_1(t)$ ($R^{\mathrm{qm}}_2(t)$) only before (after) passing through the avoided crossing. By virtue of Eq.~(\ref{eqn: mean nuclear position}), this will coincide with the trajectory followed by the mean nuclear position.

\begin{figure}[h!]
 \begin{center}
  \includegraphics*[angle=270,width=.6\textwidth]{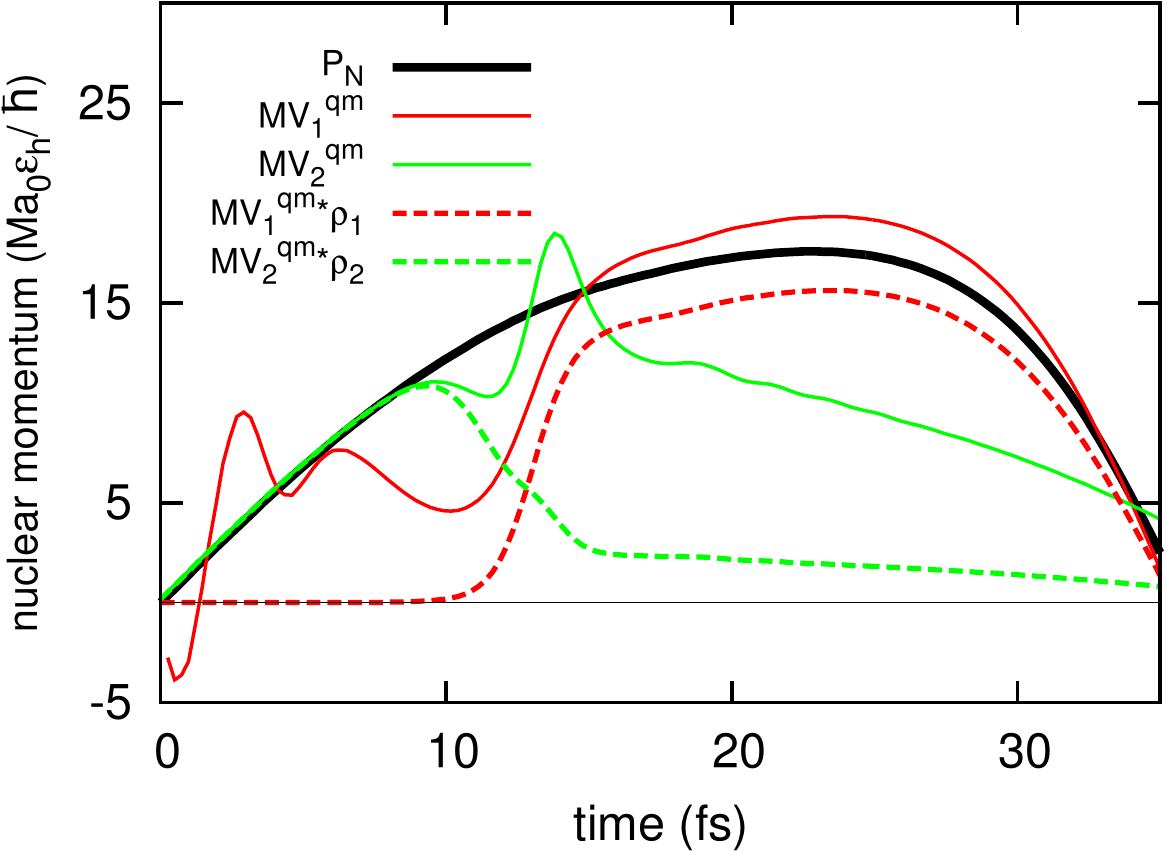}
 \end{center}
 \caption{$P_n$ (black line), $MV^{\mathrm{qm}}_j(t)$ ($j=1$ red line, $j=2$ green line) and $MV^{\mathrm{qm}}_j(t)\rho_j(t)$ ($j=1$ dashed red line, $j=2$ dashed green line) as functions of time. The thin horizontal line shows the value 0.}
 \label{fig: momenta} 
\end{figure}

Observations consistent with the results for the position can be presented for momentum. Fig.~\ref{fig: momenta} shows analogous results for the mean nuclear momentum
\begin{equation}
P_n(t)=\int dR \chi^*(t) \left[-i\hbar\partial_R\chi(R,t)\right]
\end{equation}
and the velocity of the mean positions associated to the wave packets $F_j(R,t)$
\begin{equation}
M\dot R^{\mathrm{qm}}_j(t)=M V^{\mathrm{qm}}_j(t).
\end{equation}
As it is clear from the figure, large deviation of $M V^{\mathrm{qm}}_j(t)$ for $j=1,2$ is shown along the whole dynamics, however better agreement is observed when we compare $P_n(t)$ with the values $M V^{\mathrm{qm}}_j(t)$ weighted by the population of the corresponding BO states. In Fig.~\ref{fig: momenta}, at around 12~fs, a clear inversion between the dashed green and dashed red lines is observed. A relation analogous to Eq.~(\ref{eqn: mean nuclear position}) in the case of the mean position can be derived, namely
\begin{equation}\label{eqn: mean nuclear momentum}
 P_n(t) \simeq \left\lbrace
 \begin{array}{cc}
 \rho_2(t) M V^{\mathrm{qm}}_2(t), & t<10\textrm{~fs} \\
 & \\
 \rho_1(t) M V^{\mathrm{qm}}_2(t), & t>15\textrm{~fs}.
 \end{array}\right.
\end{equation}
In both cases, for the position and for the momentum, one of the two ``weighted'' contributions is almost zero before or after the passage through the coupling region (occurring at around 12~fs). Therefore, we can anticipate that, despite the deviation of the nuclear evolution from a purely~\footnote{``Purely'' here refers to conditions discussed in Section~\ref{sec: limit}.} classical behavior, we expect a good agreement between exact (quantum mechanical) mean values and approximated classical observables.

\begin{figure}
 \begin{center}
  \includegraphics*[angle=270,width=.6\textwidth]{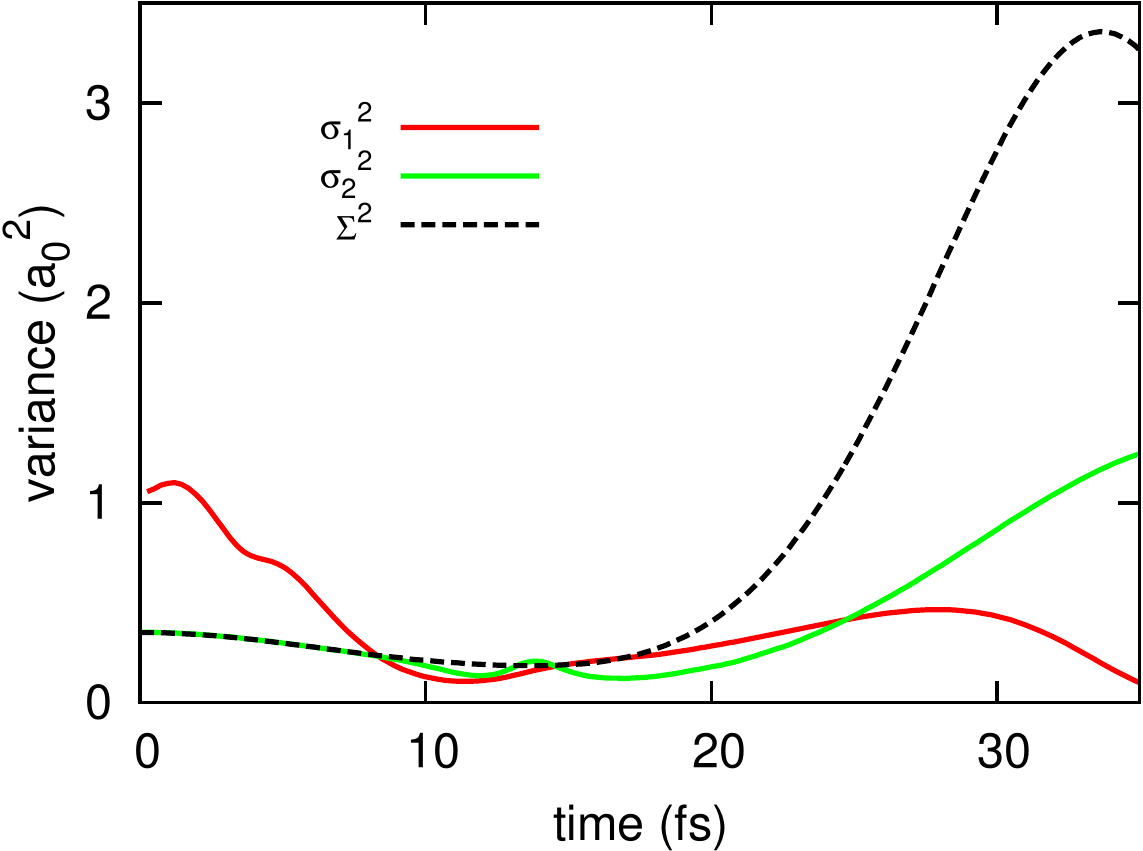}
 \end{center}
 \caption{Variances $\Sigma$ (dashed black line) and $\sigma_j$ ($j=1$ red line, $j=2$ green line) as functions of time.}
 \label{fig: sigma} 
\end{figure}

In Section~\ref{sec: limit}, it is shown that Eq.~(\ref{eqn: localization}) is valid if the variance $\Sigma$ associated to the nuclear density is equal (or close) to the variances $\sigma_j$ of $|F_j(R,t)|^2$. Therefore, we calculate the variances $\Sigma$ and $\sigma_j$ associated to $|\chi(R,t)|^2$ and $|F_j(R,t)|^2$, respectively. They are shown, as functions of time, in Fig.~\ref{fig: sigma}.
The results for the variances confirm the observations reported so far, namely after 20~fs the values largely deviate from each other. The initial disagreement between $\sigma_1^2$ and $\Sigma$ does not have a strong impact on the dynamics, since at initial times the function $|F_1(R,t)|^2$ is almost zero.

\begin{figure}
 \begin{center}
  \includegraphics*[angle=270,width=.5\textwidth]{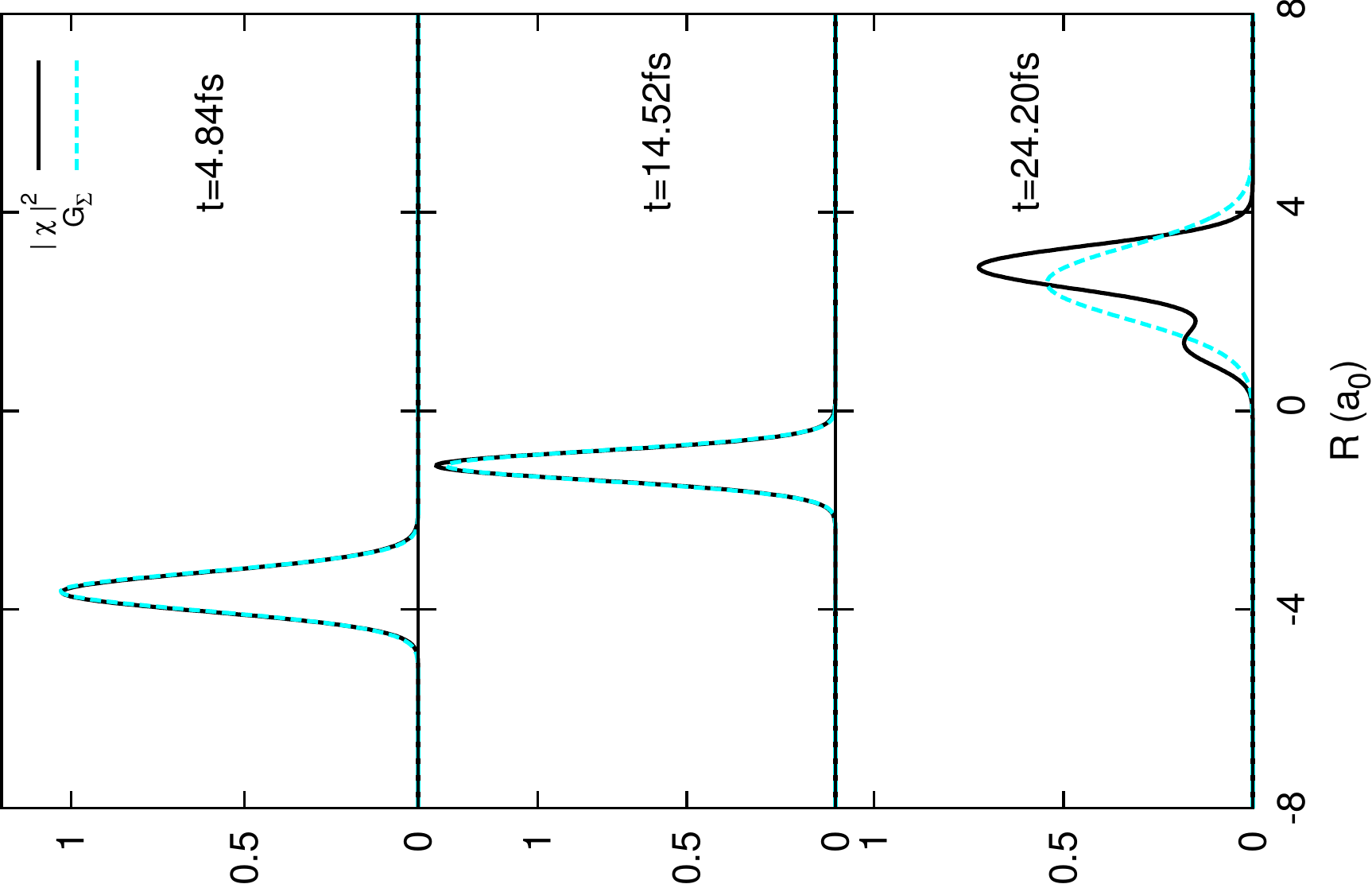}
 \end{center}
 \caption{Comparison between the nuclear density (black lines) and a Gaussian (dashed cyan lines) with the same mean position and variance at times $t=4.48,14.52,24.20$~fs.}
 \label{fig: density} 
\end{figure}

The discussion presented in Section~\ref{sec: limit} is based on the representation of the nuclear density as a Gaussian function with the aim of taking the classical limit as $\Sigma\rightarrow 0$. Therefore, it is worth showing the validity of this initial assumption. Fig.~\ref{fig: density} shows the comparison between the nuclear density at different times and a Gaussian function centered at the mean nuclear position and with variance 
\begin{equation}
\Sigma^2(t) = 2\int dR \left(R-R_n(t)\right)^2 \left|\chi(R,t)\right|^2.
\end{equation}
As expected from previous observations, the Gaussian shape of the nuclear density is lost after 20~fs and the considerations presented in Section~\ref{sec: limit} do not strictly apply after this time.

\subsection{Quantum vs. MQC evolution}\label{sec: results 2}
The electronic and nuclear equations in the MQC scheme are integrated by using the fourth-order Runge-Kutta algorithm and the velocity-Verlet algorithm, respectively, with time step $2.4\times10^{-3}$~fs ($0.1$~a.u.), as in the quantum propagation.

In the following we will show some observable computed with the MQC algorithm and we will compare approximate results with the reference exact data. Two approaches will be used, respectively referred to as single-trajectory (ST) and multiple-trajectory (MT) approaches. In the first case, a single classical trajectory is coupled to the quantum electronic evolution. In the second case, 6000 independent trajectories are employed to describe nuclear motion. In the ST approach, the classical particle is at $R_0=-4.0$~$a_0$ at the initial time with zero initial momentum, whereas in the MT case, initial positions and momenta are sampled from the Wigner distribution $\rho_W(R,P)$ associated to the initial nuclear wave packet. Since the initial wave packet is a Gaussian centered at $R_0$, the Wigner function is the product of two independent Gaussian functions, one in position space and one in momentum space, and it is positive-definite.

\begin{figure}
 \begin{center}
  \includegraphics*[angle=270,width=.6\textwidth]{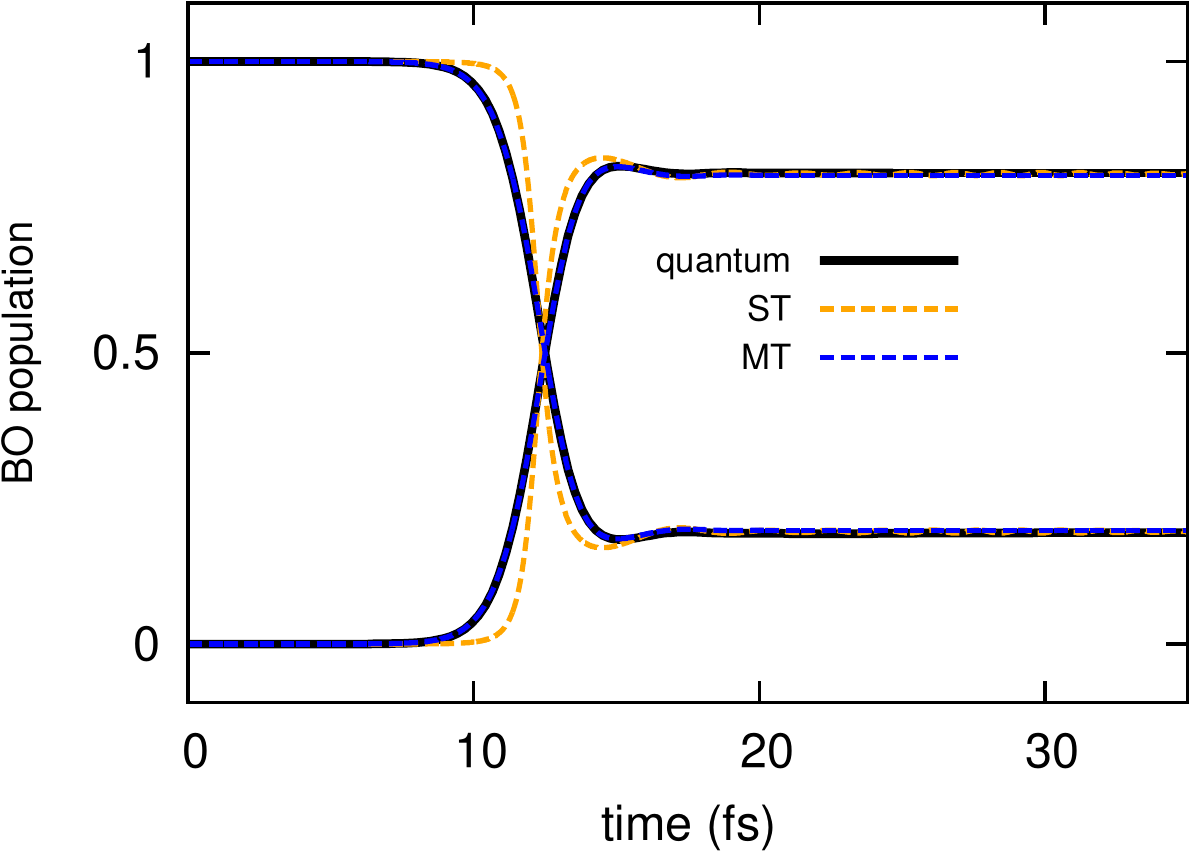}
 \end{center}
 \caption{Populations of the BO states as functions of time. Black lines represent exact (quantum) results, dashed lines are the results of the MQC approach, ST (orange) and MT (blue).}
 \label{fig: bo population} 
\end{figure}

Fig.~\ref{fig: bo population} shows the population of the BO states as functions of time. The results from quantum calculations are represented as black lines, whereas the dashed orange and dashed blue are the results of the ST and MT approximated evolution, respectively.
The branching of the electronic populations is correctly reproduced within the approximate MQC scheme, even when the ST approach is used. However, when the delocalization of the nuclear wave packet is accounted for, with the MT scheme, the agreement between reference and approximate results is perfect along the whole dynamics.

\begin{figure}[!h]
 \begin{center}
  \includegraphics*[angle=270,width=.6\textwidth]{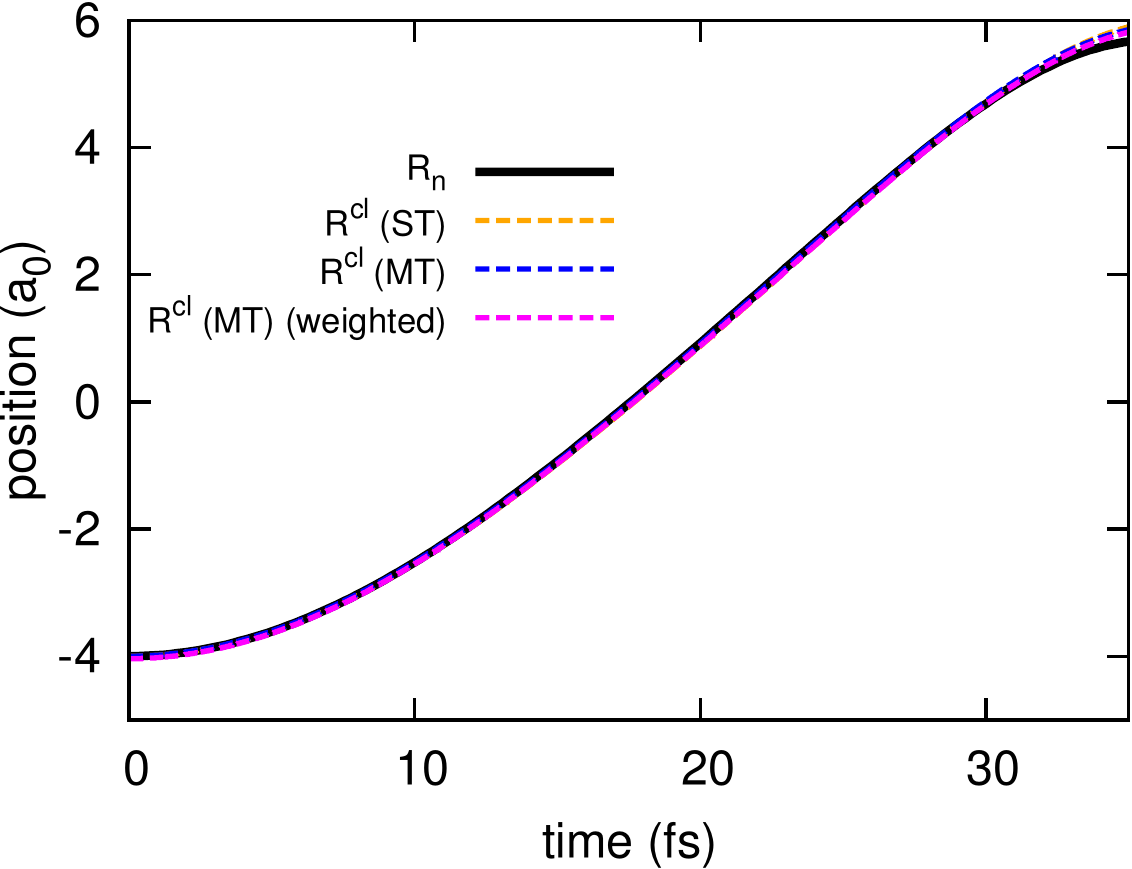}
 \end{center}
 \caption{Mean nuclear position $R_n(t)$ (black line) compared to the classical position $R^{cl}(t)$ evaluated from the ST (dashed orange line) and from the MT (dashed blue and magenta lines) approach. The dashed blue line is determined according to Eq.~(\ref{eqn: average 1}) and the dashed magenta line according to Eq.~(\ref{eqn: average 2}).}
 \label{fig: position mqc} 
\end{figure}

\begin{figure}[!h]
 \begin{center}
  \includegraphics*[angle=270,width=.6\textwidth]{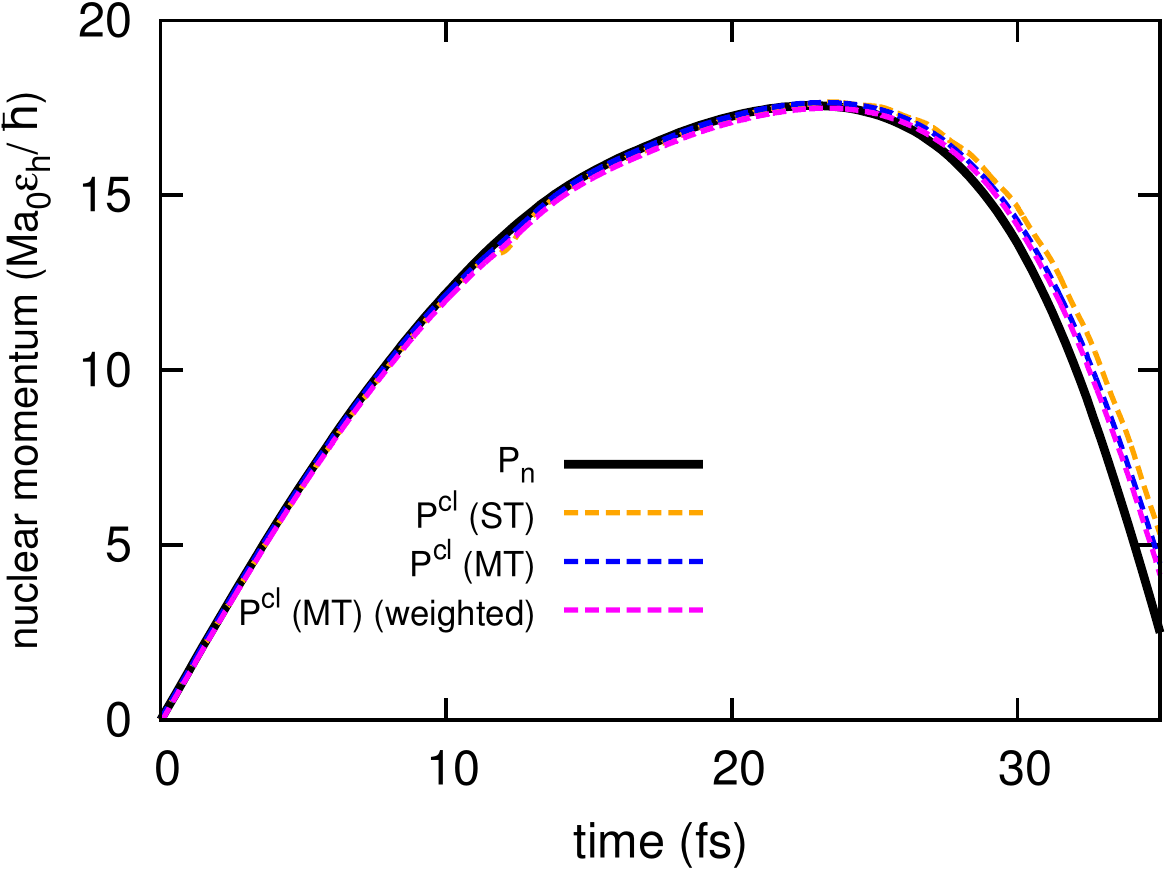}
 \end{center}
 \caption{Mean nuclear momentum. The color code is the same as in Fig.~\ref{fig: position mqc}.}
 \label{fig: momentum mqc} 
\end{figure}

\begin{figure}[!h]
 \begin{center}
  \includegraphics*[angle=270,width=.6\textwidth]{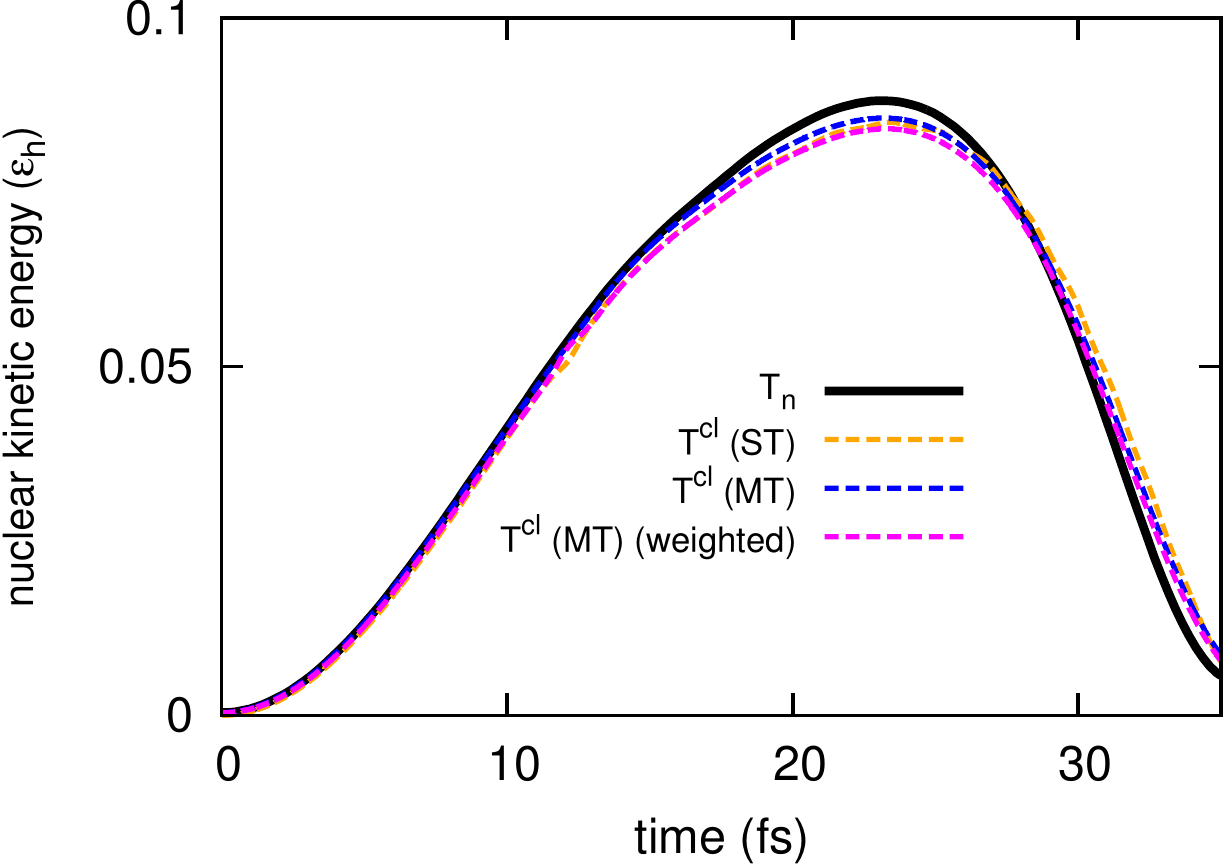}
 \end{center}
 \caption{Mean nuclear kinetic energy. The color code is the same as in Fig.~\ref{fig: position mqc}.}
 \label{fig: kinetic energy mqc} 
\end{figure}

\begin{figure}[!h]
 \begin{center}
  \includegraphics*[angle=270,width=.6\textwidth]{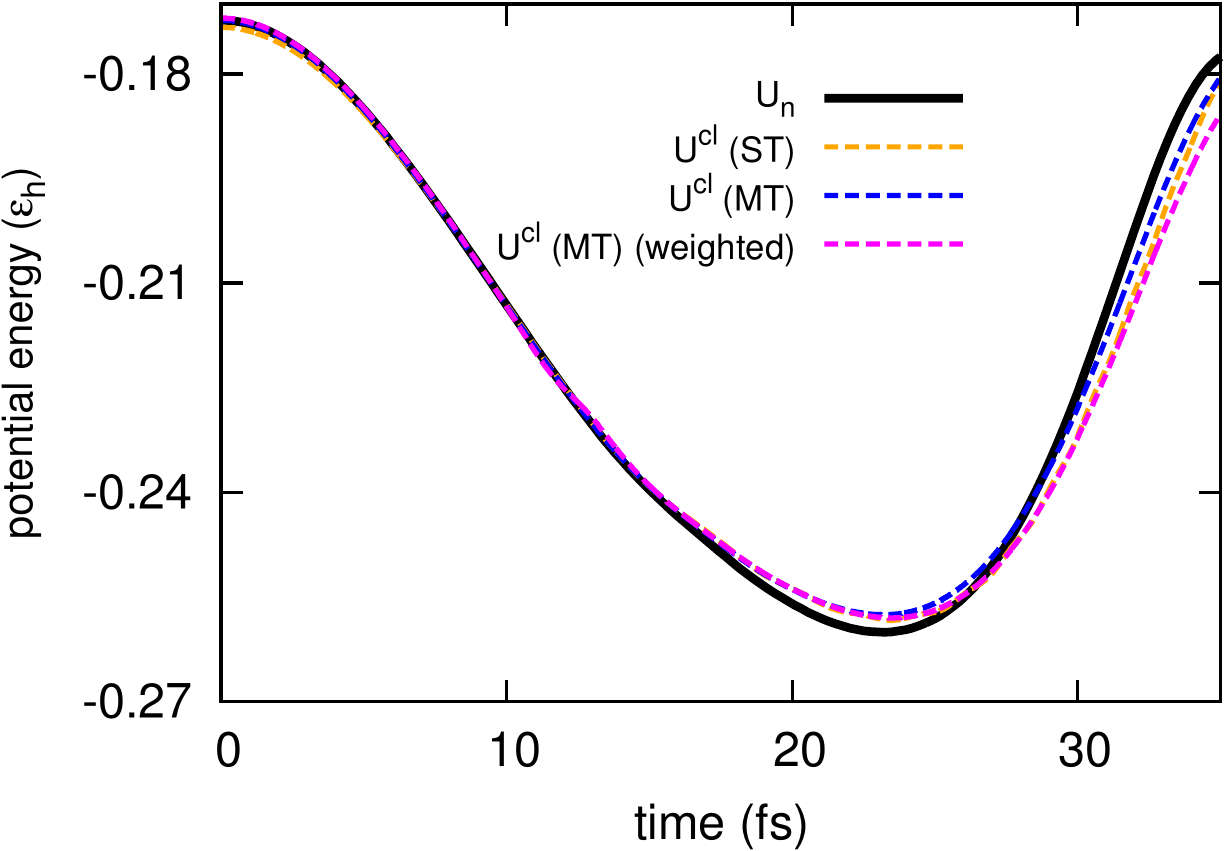}
 \end{center}
 \caption{Mean potential energy. The color code is the same as in Fig.~\ref{fig: position mqc}.}
 \label{fig: potential energy mqc} 
\end{figure}

In the following figures, we will show some nuclear observables, namely mean position in Fig.~\ref{fig: position mqc}, momentum in Fig.~\ref{fig: momentum mqc}, kinetic energy in Fig.~\ref{fig: kinetic energy mqc} and potential energy in Fig.~\ref{fig: potential energy mqc}, as functions of time.
All figures show a very good agreement between exact and MQC results, with a slight improvement of the MT approach compared to the ST approach. This is clearly expected since when using a set of independent trajectories we are taking into account the effect of the nuclear delocalization.

It is worth noting that in all plots we are using two different expressions to evaluate the mean values in the MT scheme. The dashed blue lines are calculated according to
\begin{equation}\label{eqn: average 1}
O(t)=\frac{1}{N_{traj}}\sum_{I=1}^{N_{traj}} O_I(t)
\end{equation}
where $O_I(t)$ is the instantaneous value of the observable $O$ along the $I$-th trajectory, $N_{traj}$ is the total number of trajectories and $O(t)$ is the value shown in the figures (dashed blue lines). The dashed magenta lines are determined from the expression
\begin{equation}\label{eqn: average 2}
O(t) = \int dR \,dP \,O\Big(R(t),P(t)\Big)\rho\Big(R(t),P(t)\Big)
\end{equation}
and in the figures they are referred to as ``weighted''. Here, $\rho(R(t),P(t))$ is the phase space distribution at time $t$, (a histogram) constructed from the distribution of classical positions and momenta, and $O(R(t),P(t))$ is the value (for all trajectories) of the observable $O$ at time $t$. Using this last expression to calculate the mean value of an observable allows to introduce the statistical weight of each trajectory via the classical distribution $\rho(R(t),P(t))$. In Eq.~(\ref{eqn: average 1}), the weight associated to each trajectory is the same, i.e. $1/N_{traj}$. The phase space distribution $\rho(R(t),P(t))$ is the time-evolved of the initial Wigner function $\rho_W(R,P)$ obtained from the initial nuclear wave packet. The density at time $t>0$, however, is determined according to a classical evolution equation, that indeed preserves the initial positive-definiteness of the Wigner function also at later times. This property would not hold true for a quantum propagation, as we have seen in Fig.~\ref{fig: density} that the nuclear density does not maintain a Gaussian shape.
For the sake of completeness, we present both sets of results, determined from Eqs.~(\ref{eqn: average 1}) and~(\ref{eqn: average 2}). They slightly deviate from each other and from the reference quantum results, but the general trend is the same.

Figs.~\ref{fig: position mqc} and~\ref{fig: momentum mqc} show a very good agreement between quantum and MQC results, for both ST and MT schemes. This confirms the discussion of the previous section, summarized in Eqs.~(\ref{eqn: mean nuclear position}) and~(\ref{eqn: mean nuclear momentum}).

Fig.~\ref{fig: kinetic energy mqc} shows the nuclear kinetic energy as function of time. The expectation value of the nuclear kinetic energy operator at time $t$
\begin{equation}
 T_n(t) = \int dr\,dR \,\Psi^*(r,R,t)\left[-\frac{\hbar^2\partial_R^2}{2M}\Psi(r,R,t)\right]
\end{equation}
leads to the expression
\begin{equation}\label{eqn: nuclear kinetic energy}
 \begin{split}
 T_n(t) = \frac{1}{2M}\int dR \chi^*(R,t)\left[-i\hbar\partial_R+A(R)\right]^2\chi(R,t)\\
 + \frac{\hbar^2}{2M}\int dR \left\langle\partial_R\Phi_R(t)|\partial_R\Phi_R(t)\right\rangle\left|\chi(R,t)\right|^2\\
 - \frac{1}{2M}\int dR A^2(R,t)\left|\chi(R,t)\right|^2
 \end{split}
\end{equation}
in terms of the nuclear and electronic wave functions. Therefore, the classical expression of the mean nuclear kinetic energy shown in Fig.~\ref{fig: kinetic energy mqc} is
\begin{align}
 T_n(t) &= \frac{\left[P_n(t)+A(R,t)\right]^2}{2M}  \nonumber \\
 &+\frac{\hbar^2}{2M} \left\langle\partial_R \Phi_R(t)|\partial_R\Phi_R(t)\right\rangle_r-\frac{A^2(R,t)}{2M},\label{eqn: classical nuclear kinetic energy}
\end{align}
where all $R$-dependent quantities are evaluated at the instantaneous nuclear position.
The plot of the nuclear kinetic energy shows a good agreement between MQC and quantum calculations, up to 20~fs, and after this time the two solutions slightly deviate from each other, yet keeping a satisfactory agreement. In the quantum propagation, the contribution to the nuclear wave packet propagating on the upper BO surface slows down (at $R\sim2.0$~$a_0$ the upper BO surface has a slightly positive slope), resulting in the splitting of the nuclear density, as shown at time $t=24.20$~fs Fig.~\ref{fig: density}. As discussed in Refs.~\cite{steps, long_steps, long_steps_mt}, the potential $\epsilon(R,t)$ driving the nuclear motion, i.e. the TDPES, develops some peculiar features after the nuclear wave packet passes through the avoided crossing, namely it has different slopes in different regions, being parallel to one or the other BO PES. Therefore, the BO-projected contributions to the nuclear wave packet evolve according to forces that are determined from different adiabatic PESs. When these PESs have opposite slopes, as in the case presented here, the BO-projected wave packets can move in opposite directions. The result is the deviation of the nuclear density from a Gaussian. The conditions illustrated in Section~\ref{sec: limit} for the validity of the classical limit are thus not fulfilled and the classical trajectory deviates from the quantum path.

Fig.~\ref{fig: potential energy mqc} shows the potential energy, averaged over the nuclear density, as function of time. The total nuclear energy at time $t$ is $T_n(t)+U_n(t)$ where $T_n(t)$ is given by Eq.~(\ref{eqn: nuclear kinetic energy}) and the potential contribution is 
\begin{equation}
U_n(t) = \int dR\,\left\langle \Phi_R(t)\right|\hat H_{BO}\left|\Phi_R(t)\right\rangle_r\left|\chi(R,t)\right|^2.
\end{equation}
This is the quantity shown as black line in Fig.~\ref{fig: potential energy mqc} along with the corresponding classical expressions, namely by considering $\left|\chi(R,t)\right|^2\rightarrow \delta(R-R^{cl}(t))$ in dashed orange and from Eqs.~(\ref{eqn: average 1}) and~(\ref{eqn: average 2}) in dashed blue and magenta.

\begin{figure}
 \begin{center}
  \includegraphics*[angle=270,width=.5\textwidth]{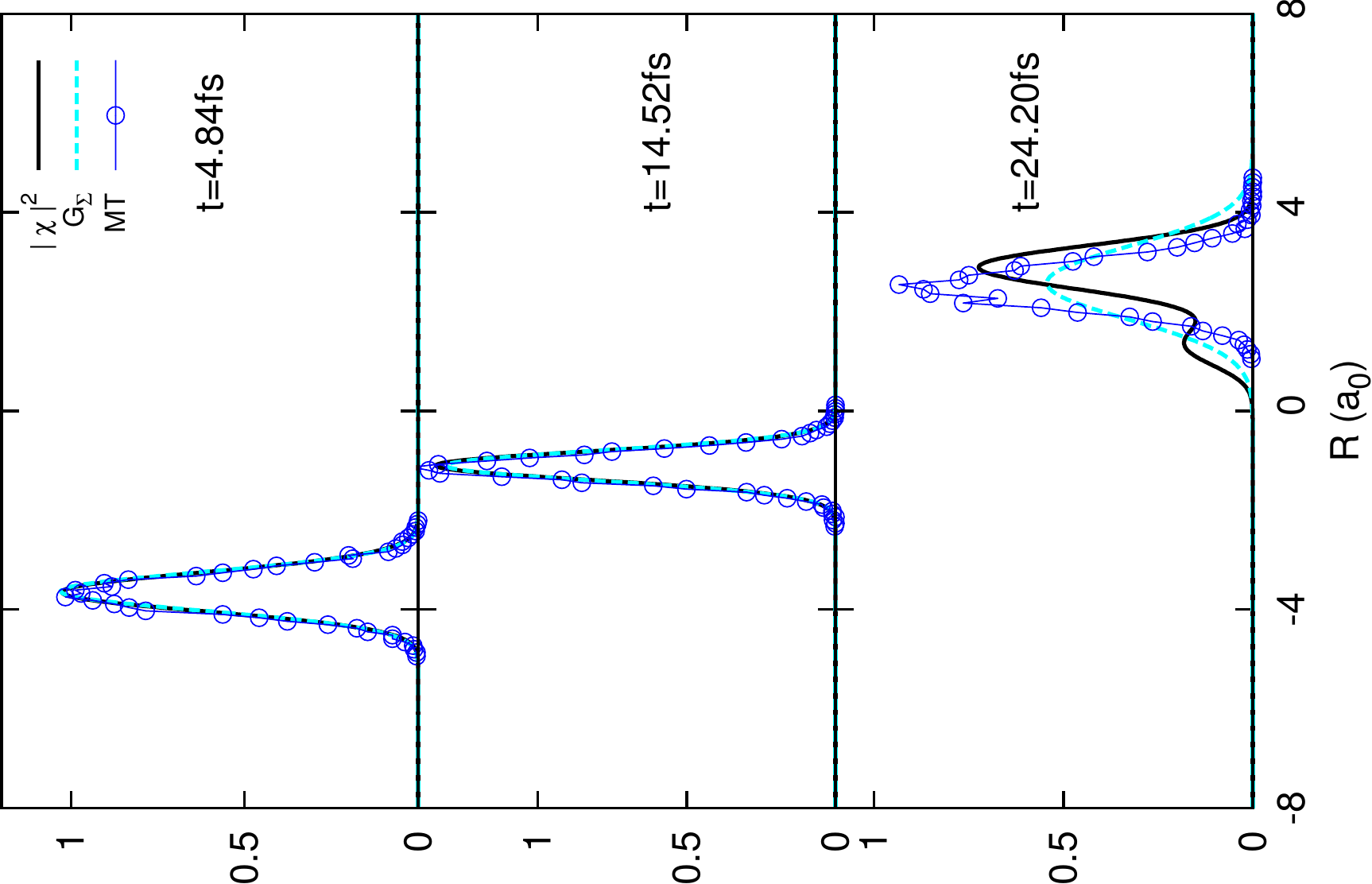}
 \end{center}
 \caption{Same as in Fig.~\ref{fig: density} by including the nuclear density from the MT scheme (blue dots), where the histogram is constructed from the distribution of classical positions.}
 \label{fig: density mqc} 
\end{figure}

As a concluding observation, we present Fig.~\ref{fig: density mqc}, which is similar to Fig.~\ref{fig: density} but we include here also the results from the MQC procedure. The nuclear density, from the expression
\begin{equation}\label{eqn: nuclear density from wigner 1}
f_{cl}\Big(R(t)\Big) = \int dP\,\rho\Big(R(t),P(t)\Big),
\end{equation}
shown as blue dots in the figure, is a histogram constructed from the distribution of classical positions. We observe that this density does not develop a double-peak structure as expected from the comparison with quantum results. This is the main cause of disagreement between the approximate scheme and the exact calculations. 

\subsection{Total and nuclear energy in the factorization framework}
The MQC method constructed here does not require to impose energy conservation along the classical trajectory. Since the method is derived from an exact approach, the conserved quantity is the total energy of the electron-nuclear system (as long as there is no external driving force)
\begin{equation}
E=\int d\dulr d\dulR \, \Psi^*(\dulr,\dulR,t) \left[\hat T_n+\hat H_{BO}\right]\Psi(\dulr,\dulR,t).
\end{equation}
This expression can be reformulated in terms of the electronic $\Phi_{\dulR}(\dulr,t)$ and nuclear $\chi(\dulR,t)$ wave functions, by introducing the time-dependent vector potential from Eq.~(\ref{eqn: vector potential}) and the (gauge-invariant part of the) TDPES from Eq.~(\ref{eqn: tdpes}). Therefore, the conserved energy is
\begin{align}
E=&\int d\dulR \,\chi^*(\dulR,t)\sum_\nu\frac{[-i\hbar\nabla_\nu+{\bA}_{\nu}(\dulR,t)]^2}{2M_\nu}\chi(\dulR,t) \nonumber \\
&+\int d\dulR\, \epsilon_{GI}(\dulR,t)\left|\chi(\dulR,t)\right|^2.
\end{align}
It is worth noting that the expression of $E$ does not contain the gauge-dependent contribution $\epsilon_{GD}(\dulR,t)=\langle \Phi_\dulR(t)|-i\hbar\partial_t|\Phi_{\dulR}(t)\rangle_{\dulr}$ to the TDPES.

In the MQC scheme derived in this paper, the expressions of the vector and scalar potentials have been approximated by neglecting all contributions arising from the spatial derivatives of the coefficients $C_j(\dulR,t)$ in Eq.~(\ref{eqn: BO expansion}). The neglected terms are the cause of deviations of the total energy from the expected constant value. However, as long as the requirements consistent with classical limit, derived in Section~\ref{sec: limit}, are satisfied, energy conservation is expected. When the classical approximation is not valid, then the energy determined using the MQC approach is not conserved and the MQC results deviate from the exact dynamics. 

\section{Conclusion}\label{sec: conclusion}
We have demonstrated that the decomposition of electronic and nuclear motion, based on the exact factorization of the molecular wave function, offers a convenient starting point for the development of a practical scheme to solve the TDSE within the QC approximation. The scheme presented here is based on the classical approximation of nuclear dynamics. In the paper we have derived the set of operations that defines the \textsl{classical limit} of the nuclear degrees of freedom. For consistency, also some properties of the electronic subsystem have been affected by this limit.

The proposed MQC method is intended to apply to those situations where quantum effects, as tunnelling or splitting of the nuclear wave packet, are not important. The numerical test presented here shows that, indeed when the behavior of the nuclei is not strictly classical, some dynamical details cannot be reproduced in the approximate scheme. However, the remarkable agreement between exact and MQC results is the evidence that the method is a promising resource for describing time-dependent processes in molecular systems involving the coupled non-adiabatic dynamics of nuclei and electrons. Further tests and improvements of the method are envisaged to introduce semi-classical corrections, in order to reproduce effects such as interference and splitting.

\section*{Acknowledgements}
Partial support from the Deutsche Forschungsgemeinschaft (SFB 762) and from the European Commission (FP7-NMP-CRONOS) is gratefully acknowledged.

\appendix
\section{The classical nuclear velocity}\label{app: classical H}
In Section~\ref{sec: limit} we introduced the following expression for the nuclear wave function
\begin{equation}\label{eqn: semiclassical chi in app}
 \chi(\dulR,t)=\exp{\left[\frac{i}{\hbar}\mathcal S(\dulR,t)\right]}
\end{equation}
by assuming that the complex function $\mathcal S(\dulR,t)$ can be expanded as an asymptotic series in powers of $\hbar$. With this hypothesis, we have shown that, at the lowest order in this expansion, classical dynamics is recovered from the TDSE by identifying $S_0(\dulR,t)$ with the classical (real) action. If the further order of the series is considered, i.e. $\mathcal O(\hbar)$, the following expression is obtained
\begin{align}
 \dot G(\dulR,t) = -\sum_{\nu=1}^{N_n}\left[\frac{\nabla_{\nu} S_0(\dulR,t)+\bA_\nu(\dulR,t)}{M_\nu}\cdot\nabla_{\nu} G(\dulR,t) \right.\nonumber \\
\left.+\frac{\nabla_{\nu}^2S_0(\dulR,t)+\nabla_{\nu}\cdot \bA_\nu(\dulR,t)}{2M_\nu}G(\dulR,t)\right],\label{eqn: continuity equation}
\end{align}
for the evolution of the function 
\begin{equation}
G(\dulR,t) = \exp{\left[i S_1(\dulR,t)\right]}.
\end{equation}
If we identify the momentum as
\begin{equation}
 \widetilde{\bf P}_\nu(\dulR,t) = \nabla_{\nu} S_0(\dulR,t)+\bA_\nu(\dulR,t),
\end{equation}
and if $G(\dulR,t)$ is a real function, then Eq.~(\ref{eqn: continuity equation}) can be written as
\begin{equation}
\partial_tG^2(\dulR,t)+\sum_{\nu=1}^{N_n}\nabla_\nu\cdot\left(\frac{\widetilde{\bf P}_\nu(\dulR,t)}{M_\nu}G^2(\dulR,t) \right)=0.
\end{equation}
This expression is (formally) a continuity equation for the function $G^2(\dulR,t)$. We identify $G^2(\dulR,t)$ as the (lowest order approximation of the) nuclear density, namely we write
\begin{align}
 G^2(\dulR,t) &= \left(\pi\Sigma^2\right)^{-\frac{1}{2}}e^{-\frac{1}{\Sigma^2}\left(\dulR-\dulR^{cl}(t)\right)^2}\nonumber \\
 &=\left(\pi\Sigma^2\right)^{-\frac{1}{2}}e^{-\frac{1}{\Sigma^2}\sum_\nu\left({\bf R}_\nu-{\bf R}_\nu^{cl}(t)\right)^2}
\end{align}
and we obtain from Eq.~(\ref{eqn: continuity equation})
\begin{align}
\Big({\bf R}_\nu -{\bf R}_\nu^{cl}(t)\Big) \cdot \left[\dot{\bf R}_\nu^{cl}(t)-\frac{\widetilde{\bf P}_\nu(\dulR,t)}{M_\nu}\right] = &\nonumber \\
-\frac{\Sigma^2}{2M_\nu}\nabla_\nu\cdot\widetilde{\bf P}_\nu(\dulR,t),\quad\forall\nu.&
\end{align}
In the limit of an infinitely localized (classical) nuclear density $\Sigma\rightarrow0$, the continuity equation leads to the definition of the classical canonical momentum
\begin{equation}
 M_\nu\dot{\bf R}_\nu^{cl}(t) = \widetilde{\bf P}_\nu(\dulR,t)=\nabla_{\nu} S_0(\dulR,t)+\bA(\dulR,t),
\end{equation}
as the rate of variation of the mean position of an infinitely localized Gaussian.

\addcontentsline{toc}{section}{References}

\end{document}